\documentclass{aa}
\usepackage{graphicx}
\usepackage{natbib}
\usepackage{color}
\usepackage{multirow}
\bibpunct{(}{)}{;}{a}{}{,}

\begin{document}

\title{On the nature of WO stars: a quantitative analysis of the WO3 star DR1 in IC~1613\thanks{Based on observations obtained at the European Southern Observatory under GTO program 088.D-0141.}}
\titlerunning{A quantitative analysis of the WO3 star DR1}
\author{F. Tramper \inst{\ref{api}} 
	\and G. Gr\"afener\inst{\ref{armagh}}
	\and O. E. Hartoog \inst{\ref{api}}
	\and H. Sana \inst{\ref{api}}
	\and A. de Koter \inst{\ref{api},\ref{leuven}}
	\and J. S. Vink \inst{\ref{armagh}}
	\and L.E. Ellerbroek \inst{\ref{api}}
	\and N. Langer \inst{\ref{bonn}}		
	\and M. Garcia \inst{\ref{madrid}}
	\and L. Kaper \inst{\ref{api}}
	\and S.E. de Mink\inst{\ref{pasadena},\ref{cit},}\thanks{Einstein Fellow}}
\institute{Astronomical Institute `Anton Pannekoek', University of Amsterdam, Science Park 904, PO Box 94249, 1090 GE Amsterdam, The Netherlands, \email{F.Tramper@uva.nl}\label{api}
	\and Armagh Observatory, College Hill, BT61 9DG Armagh, UK\label{armagh}
	\and Instituut voor Sterrenkunde, Universiteit Leuven, Celestijnenlaan 200 D, 3001 Leuven, Belgium\label{leuven}
	\and Argelander Institut f\"ur Astronomie, University of Bonn, Auf dem H\"ugel 71, D-53121, Bonn, Germany\label{bonn}
	\and Centro de Astrobiologia, CSIC-INTA, Ctra. Torrejon a Ajalvir km.4, 28850-Madrid, Spain \label{madrid}
	\and Observatories of the Carnegie Institution for Science, 813 Santa Barbara St, Pasadena, CA 91101, USA \label{pasadena}
	\and Cahill Center for Astrophysics, California Institute for Technology, Pasadena, CA 91125, USA\label{cit}}

\abstract
{Oxygen sequence Wolf-Rayet (WO) stars are thought to represent the final evolutionary stage of the most massive stars. The characteristic strong \ion{O}{vi} emission possibly originates from an enhanced oxygen abundance in the stellar wind. Alternatively, the \ion{O}{vi} emission can be caused by the high temperature of these stars, in which case the WO stars are the high-temperature extension of the more common carbon sequence Wolf-Rayet (WC) stars.}
{By constraining the physical properties and evolutionary status of DR1, a WO star in the low-metallicity Local Group dwarf galaxy IC~1613 and one of only two objects of its class known in a SMC-like metallicity environment, we aim to investigate the nature of WO stars and their evolutionary connection with WC stars.}
{We use the non-LTE atmosphere code {\sc cmfgen}\ to model the observed spectrum of DR1 and to derive its stellar and wind parameters. We compare our values with other studies of WC and WO stars, as well as with the predictions of evolutionary models. We also model the surrounding nebula using the photo-ionization code {\sc cloudy}.}
{The oxygen and carbon abundances that we obtain are comparable to values found for WC stars. The temperature and luminosity are, however, higher than those of WC stars. DR1 is embedded in the hottest known \ion{H}{ii} region in the Local Group. The nebular properties can be consistently reproduced by {\sc cloudy}\ models adopting DR1 as central ionizing source.}
{Comparison of the abundances and temperature of DR1 with core helium-burning models show that DR1 is currently well into the second half of helium burning. If the properties of DR1 are representative for the WO class, it would imply that WO stars are the high-temperature and high-luminosity extension of the WC stars, and do not necessarily represent a later evolutionary stage.}


\maketitle

\section{Introduction}
The oxygen sequence Wolf-Rayet (WO) stars, introduced by \cite{barlow1982}, are extremely rare. Only eight members of this class are currently known: Sand 4 (WR 102), Sand 5 (WR 142), MS4 (WR 30a) and WR 93b in the Milky Way, Sand 2 (BAT 99-123) and the recently discovered LH41-1042 \citep{neugent2012} in the Large Magellanic Cloud (LMC), Sand 1 (Sk 188) in the Small Magellanic Cloud (SMC), and DR1 in IC~1613. Their spectra are characterized by strong emission lines of highly ionized oxygen, in particular the \ion{O}{vi} 3811-34 \AA \ line with an equivalent width of up to 1700 \AA \ \citep{kingsburgh1995b}.

\par The origin of the high-excitation oxygen emission is widely attributed to the surfacing of this species during the late stages of core helium (or possibly carbon) burning \citep{barlow1982,smith1991}. Revealing the core at this late stage of evolution requires the stellar mass-loss rate in prior stages to be relatively low. In this scenario the presence of WO stars is therefore preferred in low-metallicity environments \citep{smith1991, georgy2009}, where the radiation-driven winds of their progenitors are relatively weak due to the low metal content \citep{vink2001, vink2005}. 

\par \cite{crowther1998} introduced a quantitative classification scheme for the carbon sequence Wolf-Rayet (WC) and WO stars, in which the WO stars are the high-temperature extension of the WC class. In this classification, the highly ionized oxygen emission is primarily the result of excitation effects, and a significant abundance difference with the WC stars is not necessarily implied. 

\par  WR stars may be subject to sub-photospheric inflation of their stellar envelopes, resulting in lower stellar temperatures \citep{grafener2012}. As this effect is expected to be more pronounced at high metallicity \citep{ishii1999}, WR stars in low-metallicity environments are expected to have higher stellar temperatures than those in the Galaxy.  

\par WO stars are often thought to represent the final stage in the evolution of stars initially more massive than 25 M$_{\odot}$ \citep{meynet2003}, i.e. including very massive stars that may avoid a red supergiant phase. If this is the case, WO stars offer the rare opportunity to study such stars just prior to their supernova (SN) explosions. Moreover, these SNe may be quite exotic, including helium-poor type Ic SNe, hypernovae \citep[e.g., ][]{nomoto2010}, and, if they retain a rapidly rotating core, even gamma-ray bursts \citep{georgy2009,woosley2006, yoon2012}.

\par The surface abundances of early WC and WO stars closely reflect the core abundances. Measuring these abundance can thus provide constraints on the controversial $^{12}$C$(\alpha, \gamma)^{16}$O thermonuclear reaction rate. 

\par In this work, we present a quantitative spectroscopic analysis of DR1 that allows us to constrain its physical and wind parameters, to investigate the nature of the object, and ultimately, its evolutionary stage. Located in the low-metallicity Local Group dwarf galaxy IC~1613, DR1 is one of the two WO stars known in a SMC-like metallicity environment (with $Z_{\mathrm{SMC}} = 1/5 \ Z_{\odot}$). Metallicity estimates for IC~1613 range from 1/10 $Z_{\odot}$ based on oxygen \citep[e.g.][]{bresolin2007} to 1/5 $Z_{\sun}$ based on iron \citep{tautvaisiene2007}. DR1 thus offers a unique probe of the final evolutionary stages of massive stars at low metallicity.
\par The layout of this paper is as follows. Section \ref{sec:previous} summarizes previous research on DR1. Section~\ref{sec:observations} describes the observations and data reduction. In Section~\ref{sec:analysis} we analyse the stellar and nebular spectra, and in Section~\ref{sec:discussion} we discuss DR1's properties, initial mass, evolutionary history and its eventual fate. Finally, Section~\ref{sec:conclusions} conveys our conclusions.

\begin{table}
\centering
\caption{Journal of observations.}\label{tab:observations}
\begin{tabular}{l  c c c}
\hline\hline
 HJD & $t_{\mathrm{exp}}$ & R-band Seeing\\
 {\tiny \textit{At start of obs.}}&  (s) & (\arcsec)\\
\hline \\[-8pt]
2455857.705 & 2 x 3600 & 0.6-0.7\\
2455858.588 & 3600 & 0.8-0.9\\
2455859.642 & 1800 & 0.9\\
\hline
\end{tabular}
\end{table}

\section{Literature overview}\label{sec:previous}

DR1 was discovered in 1982 by \cite{dodorico1982}. Based on spectra in the 4000-7000 \AA \ range, they classified it as a peculiar WC star or a WC + WN binary. In the same year, spectra extending below 3600 \AA \ were obtained by \cite{davidson1982}, who suggested that the star could be a member of the WO class because of the presence of strong \ion{O}{vi} $\lambda\lambda$3811-34 emission. \cite{davidson1982} derived a temperature of 100 kK for the star based on the nebular \ion{He}{ii} $\lambda$4686 flux and assuming a blackbody distribution for the ionizing radiation. Subsequent studies by \cite{armandroff1991} and \cite{garnett1991} yielded spectral types of WC4-5 and WO4, respectively. The latter authors estimated that the effective temperature $T_{\mathrm{eff}}$ should be in the range of 75 kK to 90 kK to reproduce the ionizing flux implied by the nebular H$\beta$ and \ion{He}{ii} $\lambda$4686 lines. Finally, DR1 and its surrounding nebula were intensively studied by \cite{kingsburgh1995a}, who adopted the spectral type WO3 \citep{kingsburgh1995b}. They derived $T_{\mathrm{eff}}$ = 75 kK, a stellar luminosity $L = 10^6 \ L_{\sun}$, and a terminal wind velocity $v_{\infty} = 2850 \ \mathrm{km \ s}^{-1}$. They reported number abundances of X(C) = 0.47, X(O) = 0.27 and X(He) = 0.25, in agreement with the values that they found for other WO stars \citep{kingsburgh1995b}.
\par DR1 is the ionizing source of the surrounding \ion{H}{ii} region S3 \citep[e.g.,][]{dodorico1982}, which exhibits unusually strong \ion{He}{ii} emission. \cite{kingsburgh1995a} derived an electron temperature $T_{\mathrm{e}} = 17.1 \ \mathrm{kK}$ and an oxygen abundance 12 + log(O/H) = 7.7, or $Z = 0.1\ Z_{\sun}$ for S3 \citep[with 12 + log(O/H)$_{\sun}$ = 8.69;][]{asplund2009}. This makes the surrounding \ion{H}{ii} region one of the hottest known in the Local Group.

\section{Observations \& data reduction}\label{sec:observations}
\subsection{Spectroscopy}

\begin{table}
\centering
\caption{INT/WFC UBVRI photometry.}\label{tab:photometry}
\begin{tabular}{l | r r r r r}
\hline\hline
Quantity & U & B & V & R & I \\
\hline \\[-8pt]
\textit{m} & 18.543 & 19.877 & 19.857 & 19.827 & 19.901 \\
\ \ \   $\sigma_m$ & 0.006 & 0.006 & 0.009 & 0.010 & 0.015 \\
\textit{M} & $-$5.74 & $-$4.41 & $-$4.43 & $-$4.46 & $-$4.38 \\
\hline
\end{tabular}
\tablefoot{
Rows one and two provide the apparent magnitude m and its error $\sigma_m$. The third row provides the corresponding absolute magnitude M, calculated adopting a distance of 721 kpc.
}
\end{table}

We have observed DR1 using the X-Shooter spectrograph \citep{vernet2011} on ESO's Very Large Telescope (VLT), which covers a wavelength range extending from the near-UV to the near-IR (3\,000 - 25\,000 \AA). The observations were carried out in October 2011 during dark time, as part of the NOVA program for X-Shooter guaranteed time. A total of 3.5 hours of integration on target has been obtained over three consecutive nights (see Table~\ref{tab:observations}).

\par The selected slit widths of 0.8", 0.9" and 0.9" result in a spectral resolving power of 6200, 8800, 5300 in the UVB, VIS and NIR instrument arms, respectively. For the NIR, a K-band blocking filter has been used to avoid reflection of both sky background and object photons from this band into the 10\,000 - 20\,000 \AA \ region, optimizing the signal-to-noise ratio in the latter wavelength range. 

\par To correct for instrument flexures, calibration frames have been taken before the start of each observation and after one hour of observations at the first night. Spectrophotometric standard stars have been observed at the beginning of each night to allow relative flux calibration. 

\par The data have been reduced using the X-Shooter instrument pipeline version 1.3.7 \citep{modigliani2010}. The extracted spectra were binned to 0.2 \AA \ in the UVB and VIS arms, and 0.6 \AA \ in the NIR arm. The (relative) flux-calibrated spectra for each observing block were co-added to obtain the final relative flux-calibrated spectrum.

\begin{figure*}[!ht]
  \resizebox{\hsize}{!}{\includegraphics{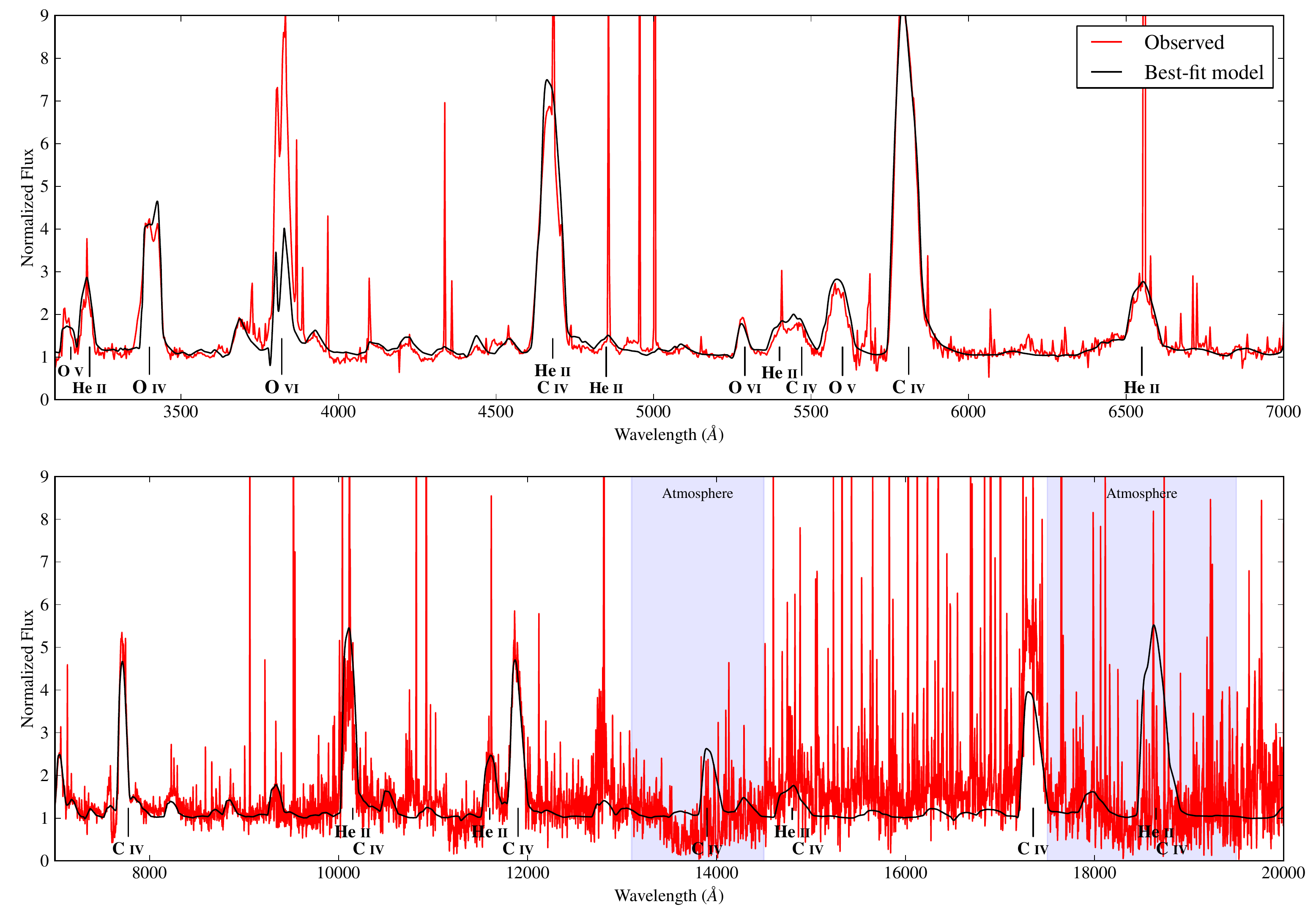}}
  \caption{Optical/near-infrared spectrum of DR1 (red). The best-fitting model spectrum is shown in black. As discussed in the text, the \ion{O}{vi} $\lambda$$\lambda$3811-34 emission is underpredicted by the model. The shaded areas indicate the regions where the atmosphere of the Earth is opaque.}
  \label{fig:bestfit}
\end{figure*}
\section{Quantitative spectroscopic analysis}\label{sec:analysis}

\subsection{Photometry}\label{sec:photometry}
The UBVRI magnitudes of the target (Table~\ref{tab:photometry}) were taken from the catalog of IC~1613's stellar population by \citet{garcia2009}. This catalog was built from PSF-photometry on multiple, dithered images of the irregular dwarf galaxy, taken with the Wide-Field Camera (WFC) at the 2.5m Isaac Newton Telescope (INT). The set of broad-band filters used, Harris-BVR and RGO-IU, are similar to the Johnson's UBVRI system. The apparent and absolute magnitudes of DR1 are presented in Table~\ref{tab:photometry}. The latter were computed adopting a foreground reddening of $E(B-V)=0.025$ \citep{schlegel1998} and a distance of 721 kpc \citep{pietrzynski2006}.

\begin{figure}[!hp]
   \resizebox{\hsize}{!}{\includegraphics{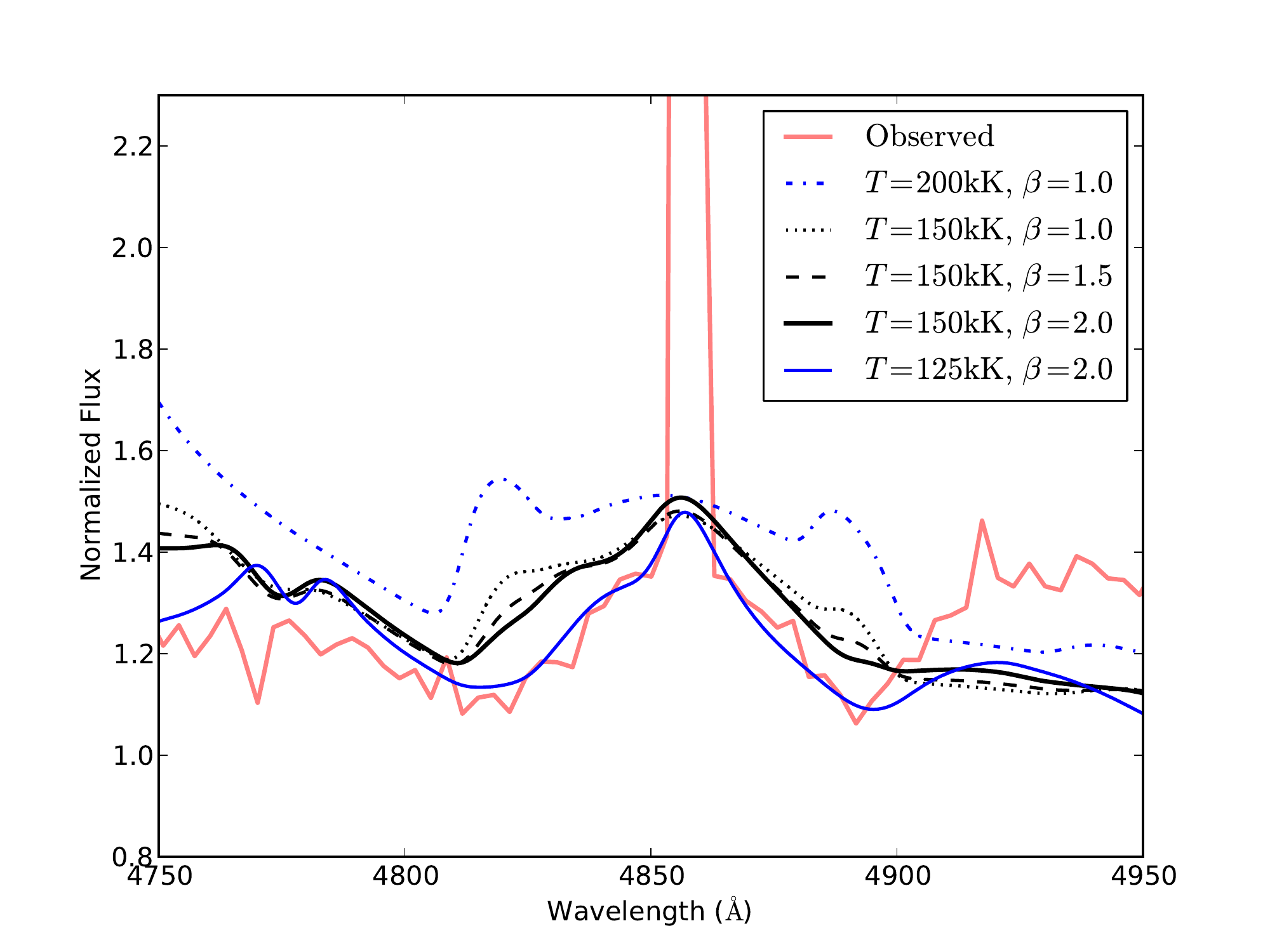}}
  \caption{Behavior of the \ion{He}{ii} line at 4859 \AA \ for different values of the temperature and $\beta$.}
  \label{fig:betaTHeII}
\end{figure}

\begin{figure}[!hp]
   \resizebox{\hsize}{!}{\includegraphics{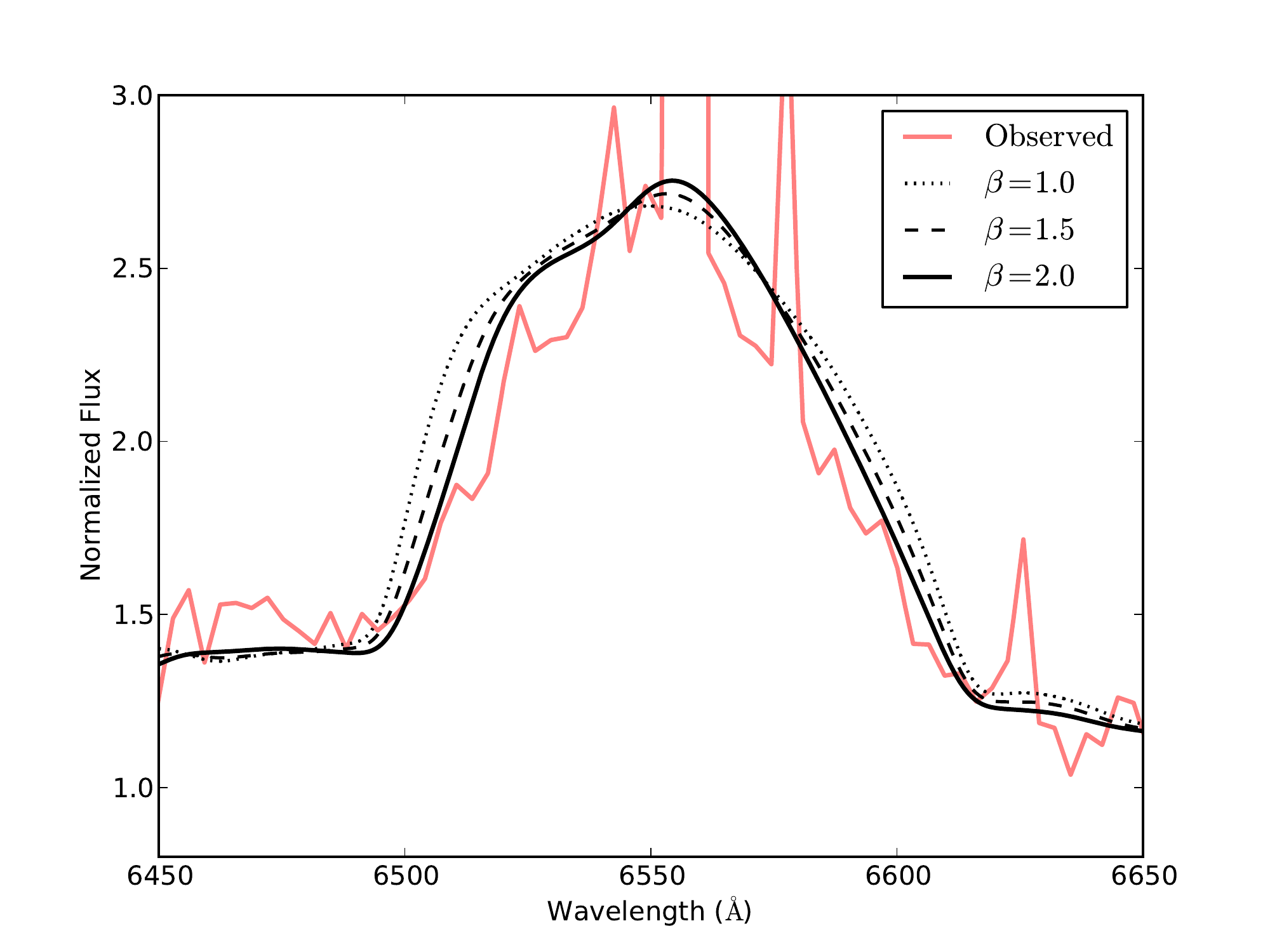}}
  \caption{Behavior of the \ion{He}{ii} $\lambda$6560 line for different values of $\beta$.}
  \label{fig:beta2}
\end{figure}

\begin{figure}[!hp]
   \resizebox{\hsize}{!}{\includegraphics{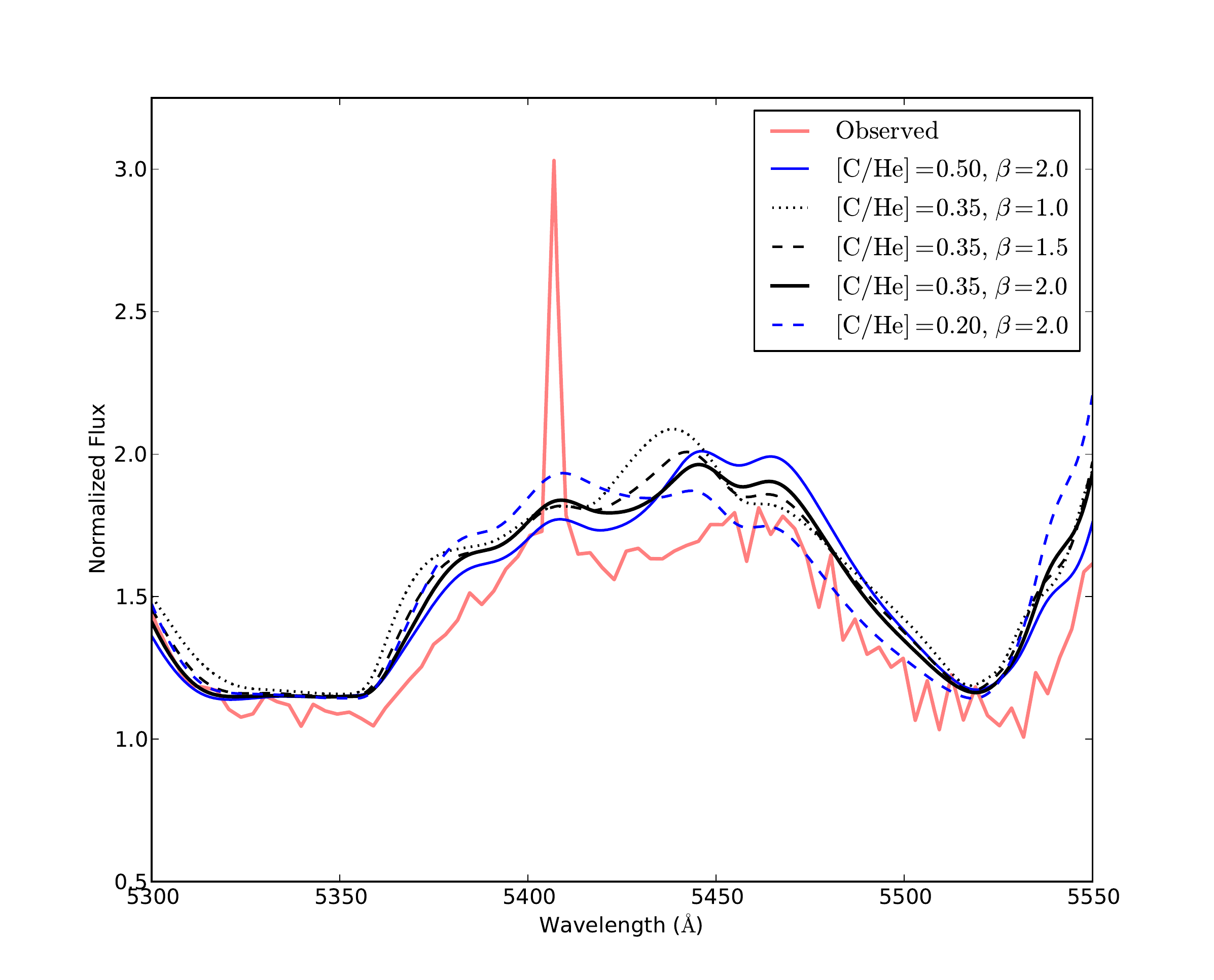}}
  \caption{Behavior of the \ion{He}{ii} and \ion{C}{iv} line complex around 5400 \AA \ for different values of $\beta$ and carbon abundance.}
  \label{fig:betacarbon}
\end{figure}

\begin{figure}[!hp]
   \resizebox{\hsize}{!}{\includegraphics{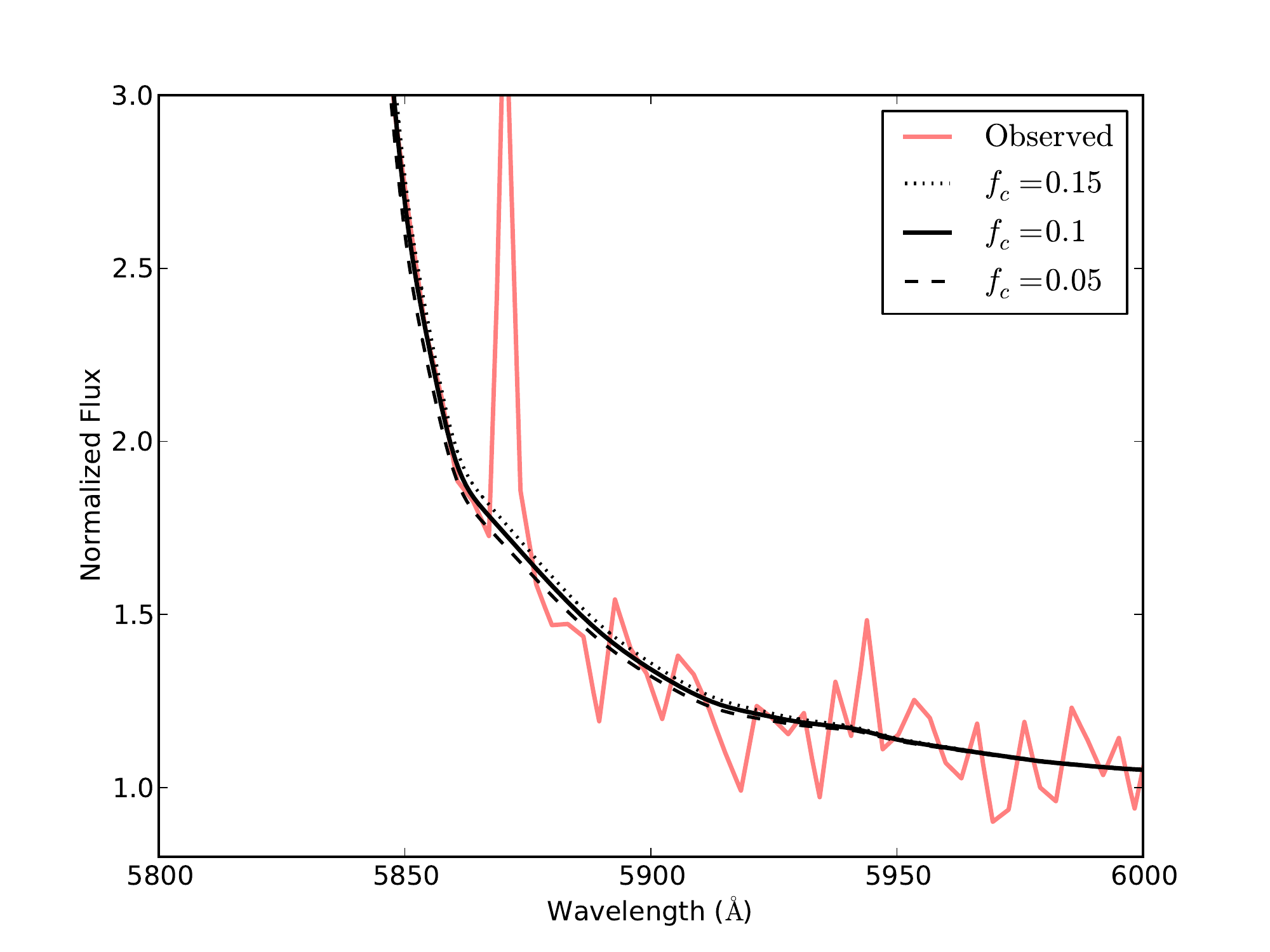}}
  \caption{Influence of the volume filling factor on the electron scattering wing of \ion{C}{iv} $\lambda$$\lambda$5801-12.}
  \label{fig:clumping}
\end{figure}

\begin{figure}[!hp]
   \resizebox{\hsize}{!}{\includegraphics{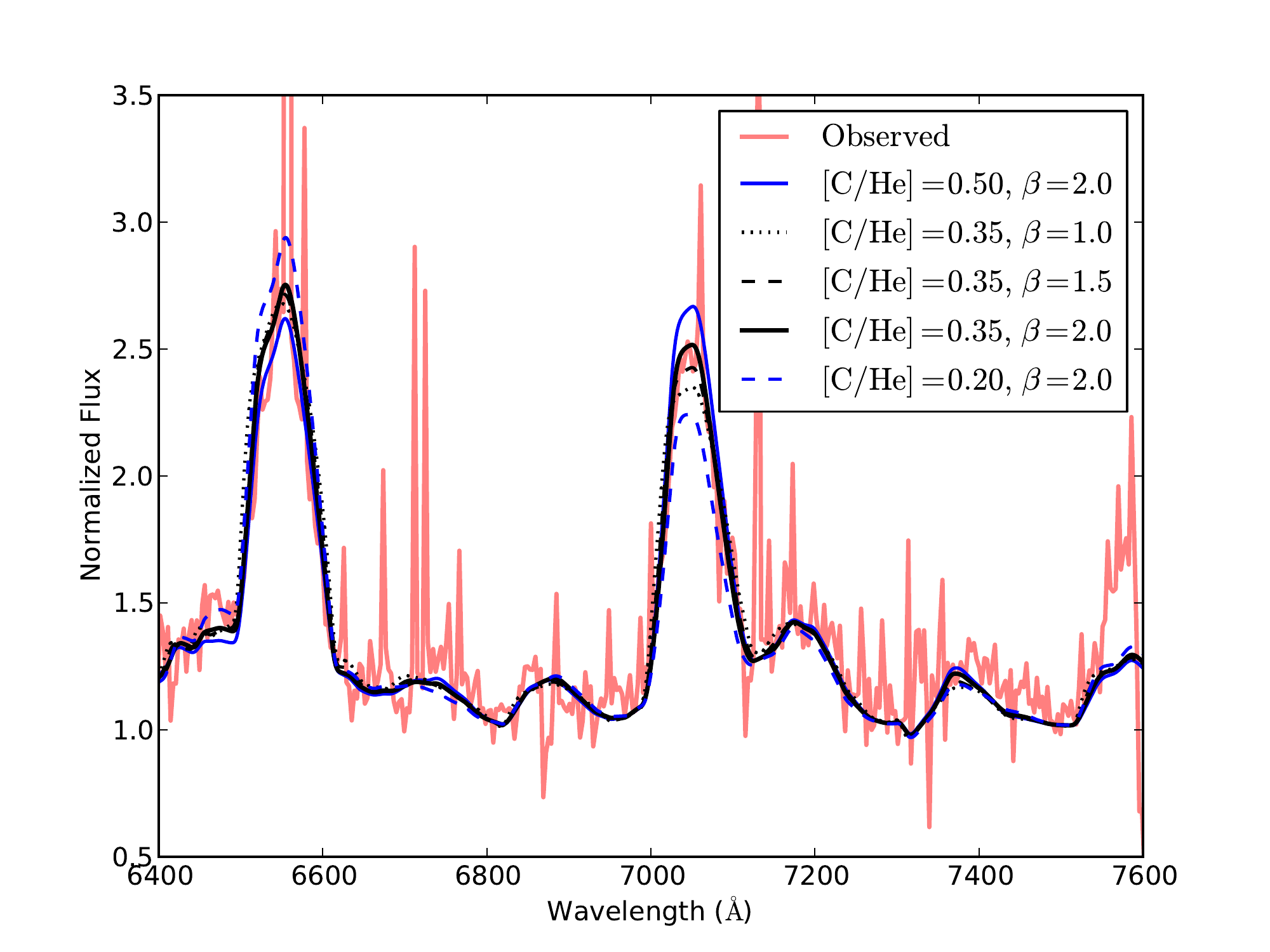}}
  \caption{Behavior of the \ion{He}{ii} $\lambda$6560 and \ion{C}{iv} $\lambda$7063 lines for different values of $\beta$ and carbon abundance.}
  \label{fig:carbon}
\end{figure}

\begin{figure}[!hp]
   \resizebox{\hsize}{!}{\includegraphics{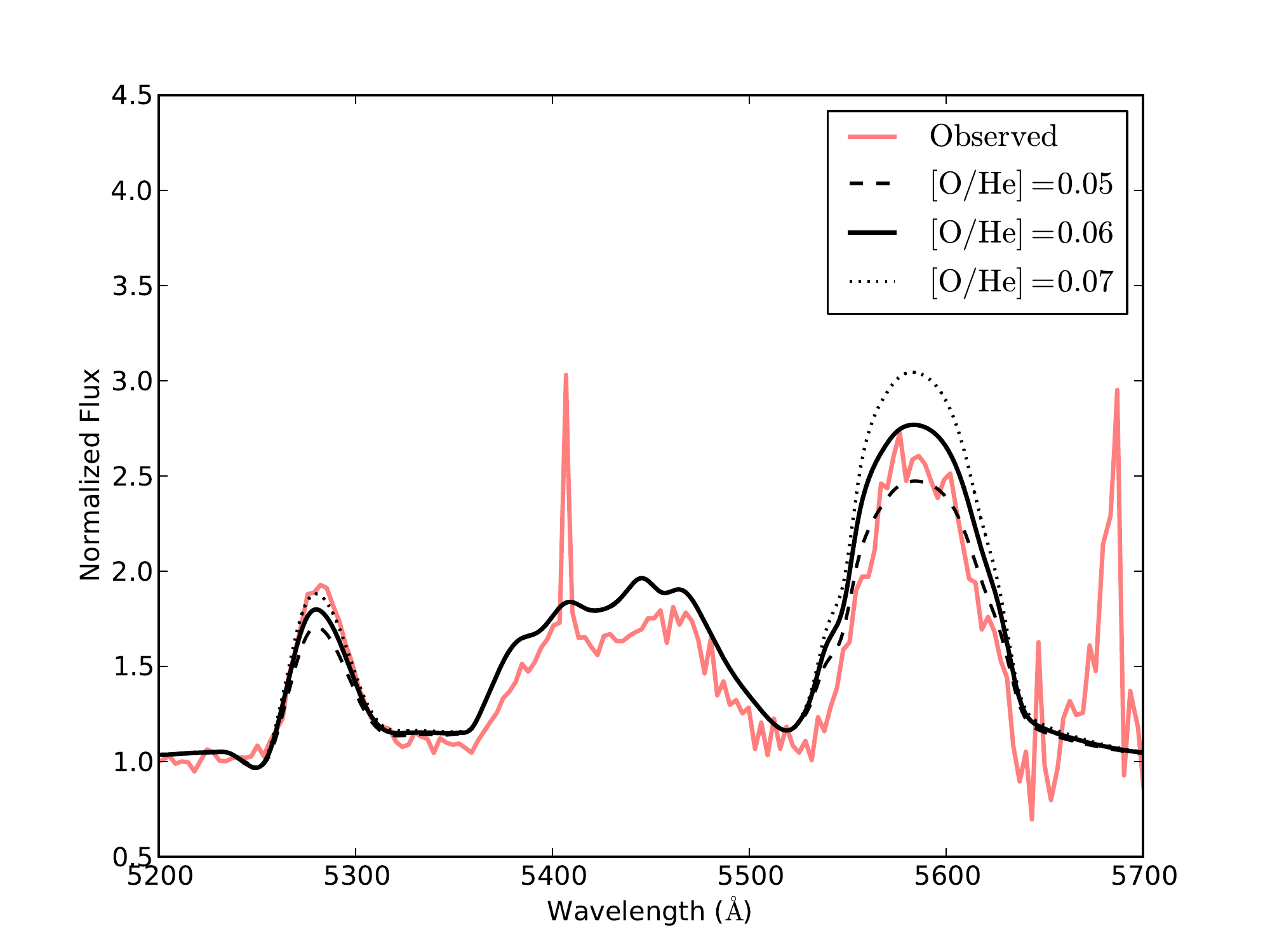}}
  \caption{Behavior of the \ion{O}{vi} $\lambda$5290 and \ion{O}{v} $\lambda$5598 lines for different values of the oxygen abundance.}
  \label{fig:oxygen}
\end{figure}

\begin{figure}[!h]
   \resizebox{\hsize}{!}{\includegraphics{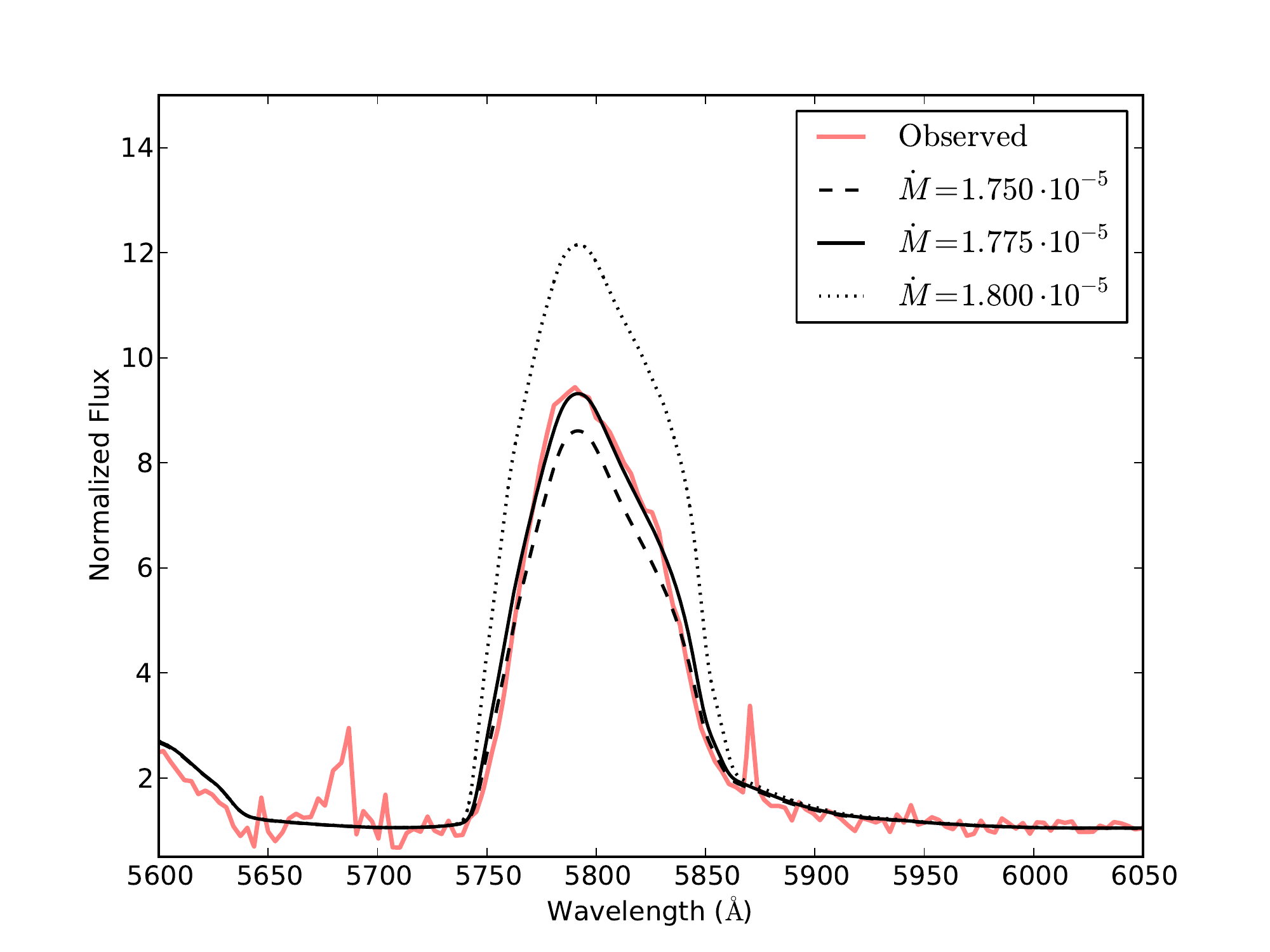}}
  \caption{Behavior of the \ion{C}{iv} line at $\lambda$$\lambda$5801-12 for different values of $\dot{M}$.}
  \label{fig:massloss}
\end{figure}

\subsection{Stellar spectrum}\label{sec:stellarspec}
We model the stellar spectrum of DR1 using the atmosphere code {\sc cmfgen} \citep{hillier1998}.  This code assumes a radial outflow of material from the atmosphere, of which the density and velocity structure is prescribed. The gas is assumed to be in a state of non-local thermodynamic equilibrium (non-LTE). The atomic models include both explicit levels and super-levels, and are of such complexity that effects such as back-warming and line-blanketing are self-consistently treated.  Convergence to a solution that fulfills radiative equilibrium can not be achieved by starting from a simple (e.g. grey LTE) solution and setting up a grid scanning the appropriate part of parameter space. Instead it requires one to migrate through a series of educated steps in specific parameters, from an existing model with fairly similar parameters to the final model.
\par This fitting procedure prevents a systematic search of parameter space and complicates an assessment of the uncertainties in the parameters of the final model. Furthermore, no specific diagnostic reacts exclusively to a given parameter, safe for luminosity which is determined from photometry. We therefore only specify error bars for this parameter.  For the other parameters we give an indication of the accuracy of the obtained value. The steps taken to arrive at the final model are discussed below.
\par In our atmosphere models, the abundances of all modeled elements except hydrogen, helium, carbon, oxygen and neon are set to a value of 10\% of solar (Table~\ref{tab:elements}). The neon abundance is enhanced by the conversion of nitrogen into $^{22}$Ne at the beginning of helium burning. The enhancement is predicted to be 1.4 times the initial oxygen mass fraction \citep{maeder1983}. For an oxygen abundance of 1/10 solar, this results in a neon mass fraction very close to the solar value, which we therefore adopt. The hydrogen abundance is set to zero. The abundances of carbon and oxygen are being fitted and are not listed in Table~\ref{tab:elements}. A summary of the ionization stages of each element included in the model is given in Table~\ref{tab:species}.
\par Although we set the metallicity of our models to $1/10 \ Z_{\odot}$, which is the metallicity based on the oxygen abundance \citep[e.g., ][]{kingsburgh1995a, bresolin2007}, there are indications that IC~1613 may have a non-solar abundance pattern. \cite{tautvaisiene2007} derived an iron abundance of $\log([\mathrm{Fe}/\mathrm{H}]) \approx -0.7$ from the analysis of three M-type supergiants, i.e. very close to the SMC iron abundance. As iron is an important driver of WR winds, we also ran a model with $Z = 1/5 \ Z_{\odot}$ to assess the impact of a higher metallicity on the derived parameters (Section~\ref{sec:metallicity}).
\par We estimate the reddening of the DR1 source to be $E(B-V) = 0.05$ by adopting the extinction law of \cite{fitzpatrick1999} and dereddening the flux-calibrated spectrum to the slope of our model continuum. We normalize the dereddened spectrum by dividing by the model continuum, and setting the flux equal to unity at 6000 \AA. The normalized spectrum does not show a noticeable slope, implying that the adopted value for the reddening is satisfactory. The best-fitting spectrum is presented in Figure~\ref{fig:bestfit}, and its parameters are listed in Table~\ref{tab:bestfit}.

\begin{table}
\centering
\caption{Mass fractions of the elements in the model.}\label{tab:elements}
\begin{tabular}{l c}
\hline\hline
 Element & Mass fraction \\
\hline \\[-8pt]
Neon & $1.74 \times 10^{-3}$ \\
Silicon & $6.99 \times 10^{-5}$ \\
Phosphorus & $6.12 \times 10^{-7}$ \\
Sulfar & $3.82 \times 10^{-5}$ \\
Chorine & $7.87 \times 10^{-7}$ \\
Argon & $1.02 \times 10^{-5}$ \\
Calcium & $6.44 \times 10^{-6}$ \\
Chromium & $1.70 \times 10^{-6}$ \\
Manganese & $9.44 \times 10^{-7}$ \\
Iron & $1.36 \times 10^{-4}$ \\
Nickel & $7.32 \times 10^{-6}$ \\
\hline
\end{tabular}
\end{table}

\subsubsection{Wind parameters: terminal velocity, $\beta$ and clumping}\label{sec:wind}

In WR spectra, the full-width at half-maximum (FWHM) of the spectral lines represents the terminal velocity of the wind ($v_{\infty}$), and this parameter can therefore easily be constrained. The best-fitting value is $v_{\infty} = 2750 \ \mathrm{km \ s}^{-1}$, and is accurate to within 50 km s$^{-1}$.
\par The wind acceleration is assumed to be described by a single-$\beta$ velocity law, i.e.
\begin{equation}
v(r) = v_{\infty} \left( 1 - \frac{R_*}{r} \right)^{\beta}.
\end{equation}
The value of $\beta$ influences the strength of the mass-loss sensitive lines (mainly \ion{O}{iv} $\lambda$3300 and \ion{C}{iv} $\lambda$5800), and is therefore degenerate with the wind strength for these lines. However, the value of $\beta$ also influences the shape of the optically thin \ion{He}{ii} lines formed at a larger radius. For low values of $\beta$ (i.e., high wind acceleration), the velocity gradient in the outer parts of the wind where these lines are formed is small, as the wind has already approached $v_{\infty}$. This causes flat-topped line profiles. For higher values of $\beta$, the wind is still accelerating towards $v_{\infty}$, and the velocity gradient at the line-forming region is larger, causing the line profiles to be more triangular-shaped (see Figure~\ref{fig:betaTHeII}). This diagnostic is somewhat degenerate with temperature (see Section~\ref{sec:temperature}), which affects the region of the wind in which these lines are formed. For higher temperatures the lines are formed further out in the wind where the velocity gradient again is smaller, thus producing flat-topped line profiles. 
\par The shape of the \ion{He}{ii} and \ion{C}{iv} line complex around 5400 \AA \ is affected by the adopted value of $\beta$. This line complex is also used to constrain the carbon abundance (see Section~\ref{sec:abundance}). The model has thus been iterated for different values of $\beta$, temperature and carbon abundance. 
\par The shape of the line profiles of \ion{He}{ii} $\lambda$4859 and  \ion{He}{ii} $\lambda$6560, as well as the shape of the \ion{He}{ii} and \ion{C}{iv} complex at 5400 \AA \ have been used to determine the best-fit value for $\beta$. Figures~\ref{fig:betaTHeII} to \ref{fig:betacarbon} show the behavior of these lines for different sets of parameters. The parameters that are not explicitly specified in these plots are conform to the numbers given in Table~\ref{tab:bestfit}. All diagnostics point toward a high value of $\beta = 2$, i.e. a slowly accelerating wind, but lower values of $\beta$ are not excluded. A high value of $\beta$ (1.5 - 2) is consistent with theoretical predictions for optically thick stellar winds \citep{vink2011}.
\par The value of the clumping parameter $f_c$ influences the shape of the electron-scattering wings of the strong lines (see Figure~\ref{fig:clumping}), although the effect is weak. The commonly adopted value of $f_c=0.1$ is in agreement with the observed spectrum, but cannot be well constrained. However, very weak clumping ($f_c \geq 0.5$) can be excluded, as \ion{O}{vi} $\lambda$5290 then gets reabsorbed in the wind and can no longer be fitted. Further constraints on the clumping are discussed in Section~\ref{sec:evolstate}.

\subsubsection{Temperature}\label{sec:temperature}
Because the stellar wind is optically thick, we do not define the effective temperature at a Rosseland optical depth of $\tau_{R} = 2/3$, as the corresponding radius is located far out in the wind. Instead, we define $T_*$ to represent the effective temperature at $\tau_R = 20$. This allows for a more meaningful comparison to evolutionary tracks, where the adopted effective temperature is not corrected for the presence of a wind.
\par Models have been calculated for temperatures $T_*$ ranging between 125 kK and 200 kK. The line ratios of the different ionization stages of carbon and oxygen do not change significantly in most  of the temperature range, and thus cannot be used to constrain the temperature. The presence of the strong \ion{O}{vi} $\lambda$$\lambda$3811-34 emission can be seen as an indication of a high temperature, although the full strength of the line cannot be reproduced. However, the line shapes of the optically thin \ion{He}{ii} lines are inconsistent with a very high temperature (see Figure~\ref{fig:betaTHeII}), as the model line profiles for these temperatures show a flat-topped shape, while the observed lines are triangular shaped (see Section~\ref{sec:wind}). 
\par A temperature of 150 kK produces the best-fitting model to all lines except \ion{O}{vi} $\lambda$$\lambda$3811-34. The underpredicted flux in this line will be further discussed in Section~\ref{sec:discussion:properties}. Although a lower temperature (125 kK) provides an even better fit to the \ion{He}{ii} lines (see Figure~\ref{fig:betaTHeII}), the line ratio of \ion{O}{vi} $\lambda$5290 to \ion{O}{v} $\lambda$5598 changes significantly for this temperature, with too much emission in the \ion{O}{v} line compared to \ion{O}{vi}. We therefore adopt 150 kK as best-fit value. Models with a temperature $T_* > 175 $ kK are excluded based on the \ion{He}{ii} line shapes (see Figure~\ref{fig:betaTHeII}).

\begin{table}
\centering
\caption{Best-fit parameters and ionizing fluxes.}\label{tab:bestfit}
\begin{tabular}{l l}
\hline\hline
Parameter & Value \\
\hline \\[-8pt]
$v_{\infty}$ & 2750 km s$^{-1}$ \\
$\beta$ & 2.0 \\
$f_c$ & 0.1 \\
$T_*$ & 150 kK \\
$[\mathrm{C/He}]$ & 0.35 \\
$[\mathrm{O/He}]$ & 0.06 \\
$\dot{M} f_c^{-0.5}$ & 5.6 $\times 10^{-5}$ $M_{\odot}$ yr$^{-1}$ \\ 
$\log{(L/L_{\odot})}$ & $5.68 {\pm 0.10}$\\
\hline \\[-8pt]
$\log{(Q_0)}$ & $49.5$ s$^{-1}$\\
$\log{(Q_1)}$ & $49.3$ s$^{-1}$\\
$\log{(Q_2)}$ & $48.0$ s$^{-1}$\\
\hline
\end{tabular}
\tablefoot{
$Q_0$, $Q_1$ and $Q_2$ are the number of ionizing photons per second for hydrogen, \ion{He}{i} and \ion{He}{ii}, respectively. Except for the luminosity, the determination of formal error bars is not possible. Uncertainties on the parameters are discussed in the text.}
\end{table}

\subsubsection{Carbon and oxygen abundances}\label{sec:abundance}
The carbon abundance (modeled as [C/He] by number), has been determined using the \ion{He}{ii} $\lambda$5412 and \ion{C}{iv} $\lambda$5471 line ratio. As can be seen from Figure~\ref{fig:betacarbon}, this diagnostic is not very sensitive to the adopted value of $\beta=2$. For this $\beta$, changes in the \ion{C}{iv} peak are minimal for different values of the carbon abundance; a value of [C/He] = 0.35 agrees best with the overall spectrum (see Figure~\ref{fig:carbon} for the typical behavior of the carbon and helium lines for different carbon abundances).
\par The oxygen abundance ([O/He] by number) can be constrained with an accuracy of 0.01 by the strength of the \ion{O}{vi} $\lambda$5290 and \ion{O}{v} $\lambda$5598 lines (see Figure~\ref{fig:oxygen}). We derive a value of [O/He] = 0.06 as best-fitting abundance.

\subsubsection{Mass-loss rate}\label{sec:massloss}
The strength of the \ion{C}{iv} $\lambda$$\lambda$5801-12 and \ion{O}{iv} $\lambda$$\lambda$3404-12 lines are very sensitive to the mass-loss rate, and therefore these lines serve as the prime diagnostic for this parameter. In general, the equivalent width of WR emission lines is found to be invariant if the transformed radius
\begin{equation}
R_t = R_* \left[ \frac{v_{\infty}}{2500 \ \mathrm{km \ s}^{-1}} \Big/ \frac{\dot{M}}{\sqrt{f_c} \ 10^{-4} \ M_{\odot} \  \mathrm{yr}^{-1}} \right]^{\frac{2}{3}}
\end{equation}
\noindent is kept fixed \citep{schmutz1989}. The derived mass-loss rate is therefore dependent on the adopted $v_{\infty}$, $f_c$, $L$ and $T_*$ (the last two values determining $R_*$).
\par For the parameters of our best-fitting model the mass-loss rate is $\dot{M} = 1.8 \times 10^{-5} \ M_{\odot} \ \mathrm{yr}^{-1}$, and can be constrained to within 0.05 dex (see Figure~\ref{fig:massloss}). The corresponding transformed radius is $R_t = 1.3 \ R_{\odot}$.

\subsubsection{Luminosity}\label{sec:luminosity}

To determine the luminosity of DR1, we have computed synthetic UVRI magnitudes of a fitted model spectrum (with $L = 3 \times 10^5 L_{\odot}$) by integrating the model flux using the transmission functions of the filters as provided by the INT website\footnote{http://catserver.ing.iac.es/filter/list.php?instrument=WFC}. Zero-point magnitudes were determined by performing spectrophotometry on a Kurucz model spectrum \citep{castelli2004} of Vega ($T_{\mathrm{eff}} = 9500$ K, $\log{g} = 4.0$, $d = 7.68$ pc, $R = 2.5 R_{\odot}$). A reddening of $E(B-V) = 0.05$ (Section~\ref{sec:stellarspec}) is then added to the obtained values. Because the \ion{O}{vi} $\lambda\lambda3811-34$ emission is underpredicted by the model, we do not use the B-band magnitude, which is affected by the flux in this line. Scaling the synthetic magnitudes to match the observed values (Table~\ref{tab:photometry}) yields a luminosity of $\log{(L/L_{\odot})}= 5.68{\pm 0.10}$. The error in this value is based on the spread in magnitude differences for each filter. The error induced by the uncertainty in the distance to IC~1613 \citep[3\%;][]{pietrzynski2006} is in comparison negligable. Note that the luminosity is overpredicted if undetected companion stars are contributing to the observed flux. After determining the luminosity, the mass-loss rate is adjusted to fit the observed line strengths.

\subsubsection{Metallicity}\label{sec:metallicity}
To determine the impact of the uncertainty in the metallicity of DR1, we computed a model with SMC metallicity ($Z = 1/5 \ Z_{\odot}$). Apart from the mass-loss rate, the derived parameters are not noticeably affected by this change. The mass-loss rate needed to fit the spectrum is slightly lower than the value derived for the low-metallicity model ($\dot{M} f_c^{-1/2} = 5.1 \times 10^{-5} \, M_{\odot} \, \mathrm{yr^{-1}}$, corresponding to $\dot{M} = 1.6 \times 10^{-5} \, M_{\odot} \, \mathrm{yr^{-1}}$ for $f_c = 0.1$). This is likely the result of more efficient line blanketing in the high-metallicity model.

\subsection{Nebular spectrum}\label{sec:nebula}
DR1 is surrounded by the ionized nebula S3. The narrow nebular emission lines are clearly distinguishable from the Wolf-Rayet spectrum (e.g., Figure~\ref{fig:bestfit}). Apart from the lines usually seen in \ion{H}{ii} regions, S3 exhibits strong \ion{He}{ii} emission (\ion{He}{ii} $\lambda$4686 /\ion{H}{$\beta$ = 0.51${\pm 0.03}$}), indicative of a hot ionizing source. The line strengths from a selection of emission lines relative to \ion{H}{$\beta$} are given in Table~\ref{tab:nebula}. The errors on the values have been calculated by the method described in \cite{hartoog2012}. Based on the X-Shooter acquisition image, we estimate the projected dimensions of the asymmetric nebula to be approximately 30 $\times$ 60 pc.
\par Following \cite{pagel1992}, we derive an electron temperature $T_{\mathrm{e}} = 17.5{\pm 0.6}$ kK based on the nebular [\ion{O}{iii}] emission. This temperature is higher than measured in other \ion{H}{ii} regions with \ion{He}{ii} emission, making S3 the hottest \ion{H}{ii} region currently known in the Local Group \citep[see][for an overview of known \ion{He}{ii} nebulae]{kehrig2011}. This electron temperature is indicative for both the high temperature of the ionizing source, and the low-metallicity environment.
\par The oxygen abundance derived from the [\ion{O}{ii}] and [\ion{O}{iii}] emission lines is $12 + \log{(\mathrm{O/H})} = 7.56{\pm 0.11}$. The electron density in the nebula is in the low density regime ($<$ 75 cm$^{-3}$) based on the [\ion{O}{ii}] $\lambda$3729/3726 and [\ion{S}{ii}] $\lambda$6716/6731 line ratios \citep{osterbrock2006}.

\begin{table}
\caption{Nebular line ratios relative to H $\beta$.}\label{tab:nebula}
\centering
\begin{tabular}{l c c}
\hline\hline
Line ID& Observed & Cloudy \\
\ \ \ \ $\lambda$(\AA)& & model \\
\hline \\[-8pt]
\ion{He}{ii} & & \\
\ \ \ \ 4686 & 0.51${\pm 0.03}$ & 0.53\\
\ion{He}{i} & & \\
\ \ \ \ 4471 & 0.017${\pm 0.007}$ & 0.022\\
\ \ \ \ 5876 & 0.058${\pm 0.014}$ & 0.058\\
\ \ \ \ 6678 & 0.017${\pm 0.007}$ & 0.016\\
$[$\ion{O}{ii}$]$ & & \\
\ \ \ \ 3726 & 0.12${\pm 0.03}$ & 0.11\\
\ \ \ \ 3729 & 0.18${\pm 0.04}$ & 0.16\\
$[$\ion{O}{iii}$]$ & & \\
\ \ \ \ 4363 & 0.13${\pm 0.01}$ & 0.10\\
\ \ \ \ 4959 & 1.45${\pm 0.07}$ & 1.39\\
\ \ \ \ 5007 & 4.36${\pm 0.18}$ & 4.20\\
$[$\ion{Ne}{iii}$]$ & & \\
\ \ \ \ 3869 & 0.63${\pm 0.06}$ & 0.40\\
\ \ \ \ 3968 & 0.19${\pm 0.08}$ & 0.12\\
$[$\ion{Ar}{iv}$]$ & & \\
\ \ \ \ 4711 & 0.061${\pm 0.011}$ & 0.058\\
\ \ \ \ 4740 & 0.053${\pm 0.019}$ & 0.045\\
$[$\ion{S}{ii}$]$ & & \\
\ \ \ \ 6716 & 0.052${\pm 0.020}$ & 0.028\\
\ \ \ \ 6731 & 0.038${\pm 0.016}$ & 0.020\\
$[$\ion{S}{iii}$]$ & & \\
\ \ \ \ 6312 & 0.024${\pm 0.017}$ & 0.008\\
\hline
\end{tabular}
\end{table}

\begin{table*}
\centering
\caption{Comparison of carbon and oxygen abundances measured in WC and WO stars.}\label{tab:abundance}
\begin{tabular}{l l l c c c c}
\hline\hline
Reference & Galaxy & Spectral type& \multicolumn{2}{c}{Mass fraction} & \multicolumn{2}{c}{Number abundance} \\
& & & \tiny{C} & \tiny{O} & \tiny{C/He} & \tiny{O/He} \\
\hline \\[-8pt]
This work & IC~1613 & WO3 & 0.46 & 0.10 & 0.35 & 0.06 \\
\cite{grafener1998} & LMC & six WC4 & 0.4 & 0.1-0.3 & & \\
\cite{crowther2000} & LMC & WO & & & 0.7 & 0.15 \\
\cite{crowther2002} & LMC & six WC4 & 0.2-0.4 & 0.1 & 0.1-0.35 & $\leq$ 0.06 \\
\cite{hillier1999} & MW & WC5 & & & 0.4 & 0.1 \\
\cite{demarco2000} & MW & WC8 & & & 0.15 & 0.03 \\
\cite{dessart2000} & MW & four WC & & & 0.08-0.25 & 0.02-0.1 \\
\cite{grafener2002} & MW & WC5 & 0.45 & 0.04 & & \\
\cite{crowther2006} & MW & WC9 & & & 0.2 & 0.01 \\
\cite{smartt2001} & M31 & WC6 & & & 0.1 & \\
\hline
\end{tabular}
\end{table*}

\par Assuming that DR1 is the dominant ionizing source of S3, we aim to reproduce the observed nebular properties using the photo-ionization code {\sc{Cloudy}} v8.0 \citep{ferland2013}. To do this, we model the \ion{H}{ii} region as a spherically symmetric cloud surrounding an ionizing source, for which we use the best-fit model of DR1. We derive nebular line strengths by computing the nebular spectrum in the line of sight towards the central source. In our models, the metal abundances in the cloud are set to 18 \% of the abundance pattern in the Orion nebula\footnote{The Orion nebula abundances used in {\sc Cloudy} are a subjective mean of values derived by \cite{baldwin1991}, \cite{rubin1991}, \cite{osterbrock1992} and \cite{rubin1993}.}, corresponding to $Z = 0.15 Z_{\odot}$ based on oxygen. The Galactic foreground extinction is discussed in Section~\ref{sec:photometry}. There is no indication that the line of sight towards DR1 suffers from extinction in IC~1613. We therefore do not include dust grains in our model. The inner radius of the cloud $r_0$ is fixed to 0.1 pc, as varying this inner radius for reasonable values ($r_0 \leq 1$ pc) does not change the resulting parameters significantly.
\par We assume the following density profile, containing both a $r^{-2}$ dependency and a constant component $n_c$:
\begin{equation}
n(r) = n_0 \left( 1 + \frac{r - r_0}{r_\mathrm{scale}} \right)^{-2} + n_{\mathrm{c}},\label{eq:density}
\end{equation}
\noindent where $n(0) + n_c$ is the density at $r_0$ and $r_{\mathrm{scale}}$ is the scale-length. We compute a grid of models varying $n_0$, $r_{\mathrm{scale}}$, and $n_c$. The line ratios of our best model are given in Table~\ref{tab:nebula}, and agree well with the observed values. 
\par The adopted model has $n_0 = 35 \ \mathrm{cm}^{-3}$, $r_{\mathrm{scale}} = 16$ pc, and $n_c = 8 \ \mathrm{cm}^{-3}$, although small variations in these parameters also give nebular properties that match the observed values. However, models with a constant density profile (i.e. $n_0 = 0$), as well as models with $r_{\mathrm{scale}} <  5$ pc, cannot reproduce the observed properties. 
\par These line ratios in Table~\ref{tab:nebula} correspond to an electron temperature $T_{\mathrm{e}}$ = 16.1 kK and an oxygen abundance of $12 + \log{(\mathrm{O/H})} = 7.61$. This abundance is consistent with the observed value, and would correspond to a metallicity of $Z \approx 0.08 \ Z_{\odot}$ based on [\ion{O}{ii}] and [\ion{O}{iii}]. As the oxygen abundance in our model is set to a value of 15\% solar, this indicates that approximately half of the oxygen is in an ionization state higher than \ion{O}{iii}. This is confirmed by the inspection of the relative population of the oxygen ionization stages in our model. Finally, the Str\"omgren radius of the modeled \ion{H}{ii} region is 15 pc, in agreement with the size of the observed nebula. The electron temperature is lower than observed, which is the case for all models that reproduce the line ratios well. This is likely caused by the assumed abundance pattern of the metals, which is observed to differ between individual \ion{H}{ii} regions \citep[e.g.,][]{zaritsky1994}. 

\par We also investigate the sensitivity of the nebular spectrum to changes in the stellar temperature, the parameter that mostly controls the production of ionizing photons. This sensitivity turns out to be very modest. Nebular models where we varied the temperature of the ionizing source by $\Delta T_{*} = 25$\,kK can also reproduce the observed nebular properties. The results of both the nebular and stellar analysis provide a consistent picture, suggesting that DR1 is indeed the main ionizing source of S3.

\section{Discussion}\label{sec:discussion}

\subsection{Oxygen abundance and temperature}\label{sec:discussion:properties}
The prominent \ion{O}{vi} $\lambda$$\lambda$3811-34 emission is the tell-tale signature of WO stars. However, this line is notoriously difficult to reproduce by models of their atmospheres. Two modeling approaches can be followed. Either the focus is to reproduce the prominent \ion{O}{vi} $\lambda$$\lambda$3811-34 emission while accepting a poorer fit to the overall spectrum, or the aim is to reproduce the entirety of the spectrum, accepting a relatively poor fit to \ion{O}{vi} $\lambda$$\lambda$3811-34.
\par The modeling of the galactic WO stars WR102 and WR142 (both with spectral type WO2) by \cite{sander2012} focusses on reproducing the \ion{O}{vi} $\lambda$$\lambda$3811-34 emission. This can be achieved by adopting a high temperature and an oxygen abundance twice as high as they on average find for early-type WC stars. However, the model spectrum underestimates the flux in some of the other lines seen in the observed spectrum, such as the \ion{O}{v} $\lambda$5598 emission. This is likely due to the high adopted temperature, which causes the higher ionization states to be preferred. 
\par \cite{crowther2000} modeled the far-UV and optical spectrum of the LMC WO star Sand 2 taking the alternative approach, and in their model do not reproduce the \ion{O}{vi} $\lambda\lambda$3811-34 emission. Their obtained temperature is 50 kK lower than temperatures obtained by \cite{sander2012} who primarely modelled \ion{O}{vi} $\lambda\lambda$3811-34, although the difference can also be attributed to the difference in spectral type.
\par As our data cover a large wavelength range (3000-20000 \AA), we focus on obtaining a good overall fit while neglecting the \ion{O}{vi} $\lambda$$\lambda$3811-34 line. Although the strong  \ion{O}{vi} $\lambda$$\lambda$3811-34 emission is underpredicted by a factor 2-3, the strength of the \ion{O}{vi} $\lambda$5290 line is well represented by our models. This suggests the presence of an additional mechanism not accounted for in the modelling, which significantly contributes to the \ion{O}{vi} $\lambda$$\lambda$3811-34 emission. 
\par The \ion{O}{vi} $\lambda$$\lambda$3811-34 emission is particularly sensitive to temperature, oxygen abundance and wind strength. A higher temperature will increase the \ion{O}{vi} to \ion{O}{v} ionization ratio, producing stronger emission in \ion{O}{vi}. A higher oxygen abundance will increase the strength of oxygen lines relative to that of lines of other elements.  The dependence of the strength of \ion{O}{vi} $\lambda$$\lambda$3811-34 on the mass-loss rate is more subtle: as the \ion{O}{vi} $\lambda$$\lambda$3811-34 emission is formed in deep layers of the stellar wind, part of the emission is absorbed by the outer layers of the wind itself. A higher mass-loss rate, i.e. a denser wind, will therefore result in weaker observed emission of this line. 
\par A physical motivation for the poor modelling of the \ion{O}{vi} $\lambda$$\lambda$3811-34 line may be found in its susceptibility to soft X-ray emission at the base of the outflow, e.g. due to shocks in the wind-acceleration zone. Such shocks could pump the upper level through the 2p$^2$p$^0$ - 3s$^2$s transition at 184 \AA. If the X-ray production is quite localized at the base of the wind, the layers above this zone may shield (part of) this X-ray emission. This prevents the \ion{C}{iv} $\lambda$$\lambda$5801-12 line, which originates from the same electron configuration transition as \ion{O}{vi} $\lambda$$\lambda$3811-34 but is formed further out in the wind, to react in a similar way.
\par Neglecting the \ion{O}{vi} $\lambda$$\lambda$3811-34 line, the carbon and oxygen abundances that we obtain are comparable to values found for early-type WC stars in a variety of environments (see Table~\ref{tab:abundance}). We thus conclude that the carbon and oxygen abundance in DR1 present no indication of enhancement compared to WC stars.

\subsection{Evolutionary state}\label{sec:evolstate}
As most massive stars are formed in close binary systems \citep{sana2012, sana2013}, it is possible that DR1 has or has had a close companion star. If DR1 is part of a close binary, it is likely that mass transfer between the stars occurs at some point during the evolution of the system, influencing their evolution. For instance, if DR1 has transferred mass onto a companion star, less mass loss through a stellar wind is needed to expose the helium-burning products. Vice versa, if DR1 has gained mass or is the product of a merger, this will likely have led to high rotation rates \citep[e.g.,][]{deMink2013}. This may lead to enhanced mass-loss due to rotation and rotational mixing, also making it easier for the helium-burning products to surface. Because we have no indication for binarity of DR1, we limit the discussion of its evolutionary state to a single-star perspective. However, we cannot exclude the possibility that DR1 is the product of binary interaction.

\par
\begin{figure}
  \resizebox{\hsize}{!}{\includegraphics{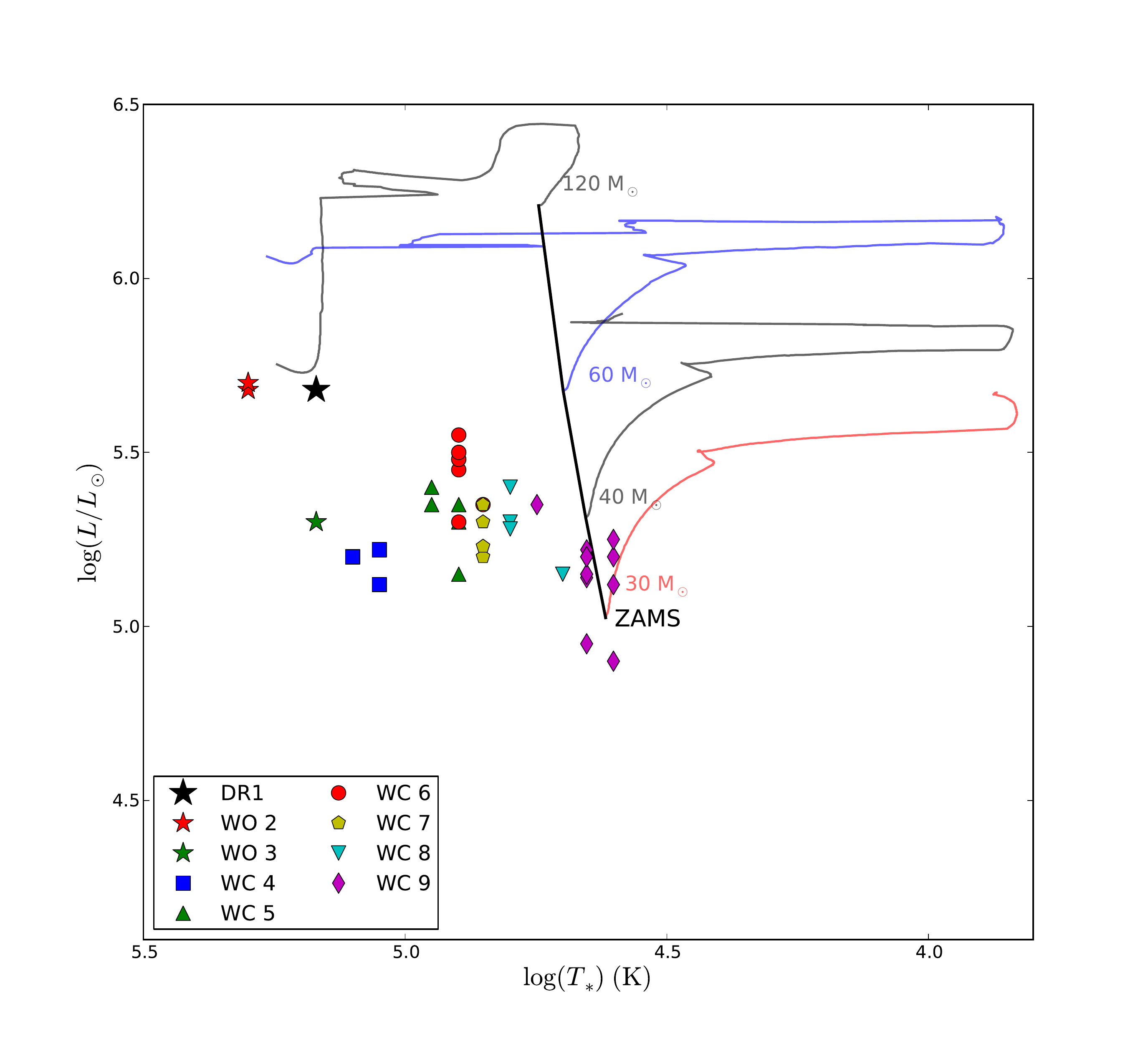}}
  \caption{Location of DR1 in the Hertzsprung-Russell Diagram. Also plotted are the WO2 and WC stars analyzed by \cite{sander2012}, the LMC WO3 star analysed by \cite{crowther2000}  and evolutionary tracks for SMC metallicity from \cite{meynet2005} for an initial rotation rate of 300 km s$^{-1}$.}
  \label{fig:HRDlowZ}
\end{figure}

\begin{figure}
  \resizebox{\hsize}{!}{\includegraphics{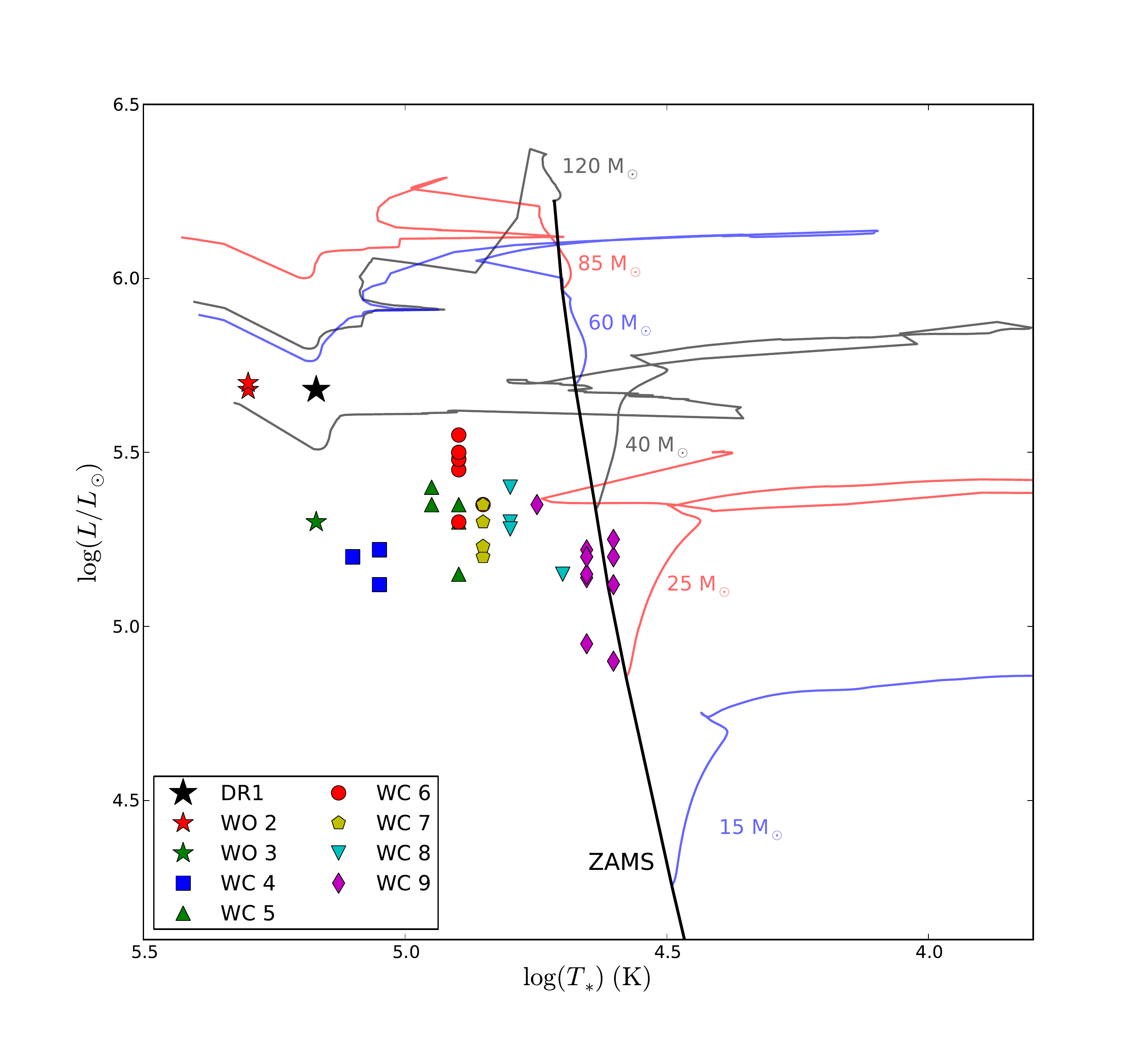}}
  \caption{Same as Figure~\ref{fig:HRDlowZ}, but with evolutionary tracks for Galactic metallicity from \cite{ekstrom2012} for an initial rotation rate of 0.4 $v_{\mathrm{crit}}$.}
  \label{fig:HRDrotation}
\end{figure}

 Figures~\ref{fig:HRDlowZ} and \ref{fig:HRDrotation} show the position of DR1 in the Hertzsprung-Russell diagram (HRD). For comparison, the two Galactic WO stars and the WC stars analyzed by \cite{sander2012}, and the WO star analysed by \cite{crowther2000} are also plotted. Both DR1 and the two WO2 stars from \cite{sander2012} are positioned at the high temperature and high luminosity regime of the strip in the HRD occupied by the WC and WO stars. The LMC WO3 star Sand 2 from \citet[][for which the luminosity is much less uncertain than the Galactic cases due to the better constrained distance]{crowther2000} is considerably less luminous, while having the same temperature as DR1. Both DR1 and Sand 2 are located very close to the helium ZAMS, indicating that they could be the descendants of stars with a different initial mass. Alternatively, it is possible that the luminosity of DR1 is overestimated if unseen companions contribute significantly to the observed flux.
\par Figures~\ref{fig:HRDlowZ} and \ref{fig:HRDrotation} also show the evolutionary tracks from \cite{meynet2005} and \cite{ekstrom2012}, representing single stars with SMC metallicity initially rotating at 300 km s$^{-1}$ and single stars with Galactic metallicity initially rotating at 40\% of critical, respectively. Figure~\ref{fig:HRDlowZ} shows that DR1 is located at a position in the HRD that is coinciding with the track for the late stages of evolution of a 120 $M_{\odot}$ star at SMC metallicity. The corresponding current-day mass would be $\approx 18 \ M_{\odot}$. This is consistent with the mass of $20 \ M_{\odot}$ that is computed from the mass-luminosity relation of \cite{grafener2011}. A firm determination of the initial mass of DR1 cannot be made, however, as the track for stars with initial masses higher than $60 \ M_{\odot}$ also reaches the high-temperature domain of the HRD. Uncertainties that may complicate an identification of the evolutionary stage of DR1 from these tracks include the initial rotational velocity, metal content and the mass-loss properties throughout the different evolutionary phases (in particular during the luminous blue variable and red supergiant phases). In general, it is thought that higher rotation rates lead to a higher mass-loss rate, and therefore to lower stellar masses during the WC or WO stage. In specific cases, however, this trend may be broken \citep[see][]{meynet2003}.
\par For a Galactic environment, Figure~\ref{fig:HRDrotation} shows that evolutionary tracks  for initial masses higher than $40 \ M_{\odot}$ reach the high-temperature domain of the HRD where the WO stars are located. Although suffering from the same uncertainties as listed above, it suggests that the occurrence of WO stars is slightly favored at higher metallicities, as the mass range for potential progenitors is larger. This is in contrast with predictions based on the assumption that the oxygen abundance is enhanced in WO stars, in which case their formation is favored in sub-Galactic metallicity environments \citep[see, e.g.,][their Figure 1]{georgy2009}.

\par
\begin{figure}
  \resizebox{\hsize}{!}{\includegraphics{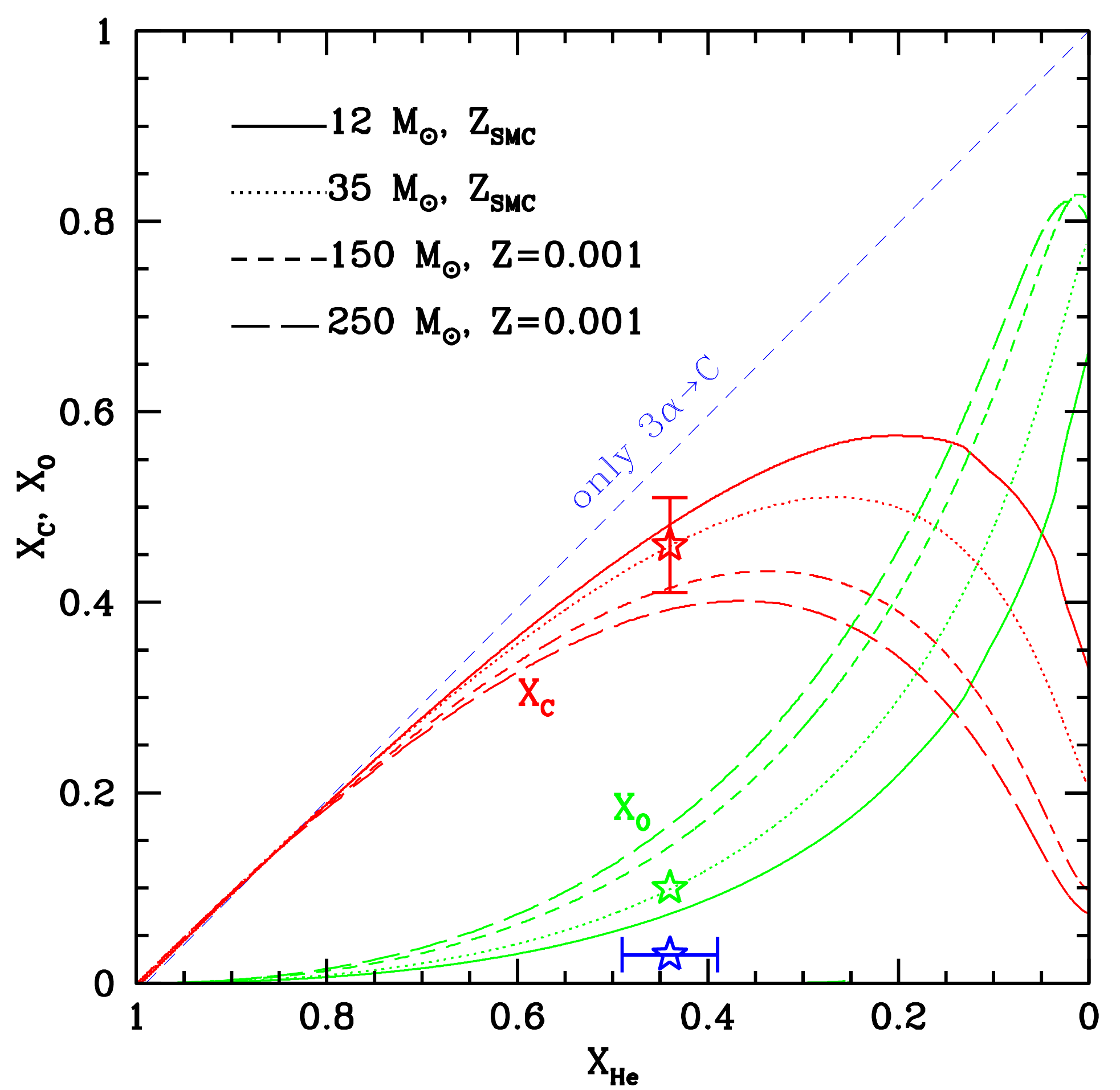}}
  \caption{Comparison of the helium (blue), carbon (red) and oxygen (green) mass fractions of DR1 with evolutionary predictions of helium-burning stars of various masses ($12\,M_{\odot}$ and $35\,M_{\odot}$ from \cite{brott2011}, $150\,M_{\odot}$ and $250\,M_{\odot}$ from \cite{langer2007}). The uncertainty of the oxygen abundance is comparable to the size of the symbol and therefore not indicated.}
  \label{fig:abundance}
\end{figure}

 Figure~\ref{fig:abundance} compares the carbon and oxygen abundances of DR1 to evolutionary predictions of the change of the helium, carbon and oxygen abundances during core-helium burning. While low-metallicity models have been used for this comparison, the influence of the metallicity is negligible. Higher masses, however, lead to a markedly lower carbon abundance and correspondingly higher oxygen abundance at a given helium mass fraction. Note that only two of the three abundances are independent, as the sum of all three is very close to one.
\par The  $^{12}$C$(\alpha, \gamma)^{16}$O thermonuclear reaction rate used in these models is still highly uncertain \citep{tur2007}. For instance, a higher $^{12}$C$(\alpha, \gamma)^{16}$O cross section would lead to somewhat lower carbon abundances at a given helium mass fraction. The reaction rate  employed in the models shown in Figure~\ref{fig:abundance} is the one proposed by \cite{weaver1993}, which appears to be required for massive stars to reproduce the Solar abundance pattern between oxygen and calcium. If this rate is correct, Figure~\ref{fig:abundance} indicates that DR1 is likely not the descendant of a star of initially several 100 $M_{\odot}$.
\par From Figure~\ref{fig:abundance}, we also see that the surface composition of DR1 corresponds to material that was in the core of the star roughly half-way into core-helium burning. This implies that DR1 must indeed be well advanced in its core-helium burning stage. Assuming a current mass of $20\,M_{\odot}$, the radiative envelope is expected to be $\approx 4.7\,M_{\odot}$ \citep{langer1989}. Based on the mass-loss rate we found for DR1, the envelope is lost at a timescale of at least $\approx 8-9\times10^4$ years (for an unclumped wind, i.e. $f_c = 1$). For a helium-burning timescale of $4.3\times10^5$ year \citep{langer1989}, we find that DR1 is currently at least three-quarters into core-helium burning. Adopting a probable post-core helium burning lifetime of $10^4$ year \citep{langer1989}, this also suggests that the volume filling factor cannot be much lower than $f_c \approx 0.2$, as otherwise the star should have already exploded. The temperature of DR1 is consistent with the temperature predicted for core-helium burning stars, while it is too low to correspond to post-core helium burning \citep{langer1988}.
\par Turning the argument around, the mass-loss rate has obviously been large enough to uncover the material which was inside the convective core in the middle of core-helium burning, i.e. {\em before} core-helium burning finished. Adopting the helium-burning timescale given above, this implies that $\dot{M} > 4.7\, M_{\odot}/ \frac{1}{2}\times4.3 \times 10^5 \, \mathrm{yr} \approx 2.2 \times 10^{-5}\, M_{\odot}\, \mathrm{yr}^{-1}$.

\subsection{Mass-loss rate}
\par
\begin{figure}
  \resizebox{\hsize}{!}{\includegraphics{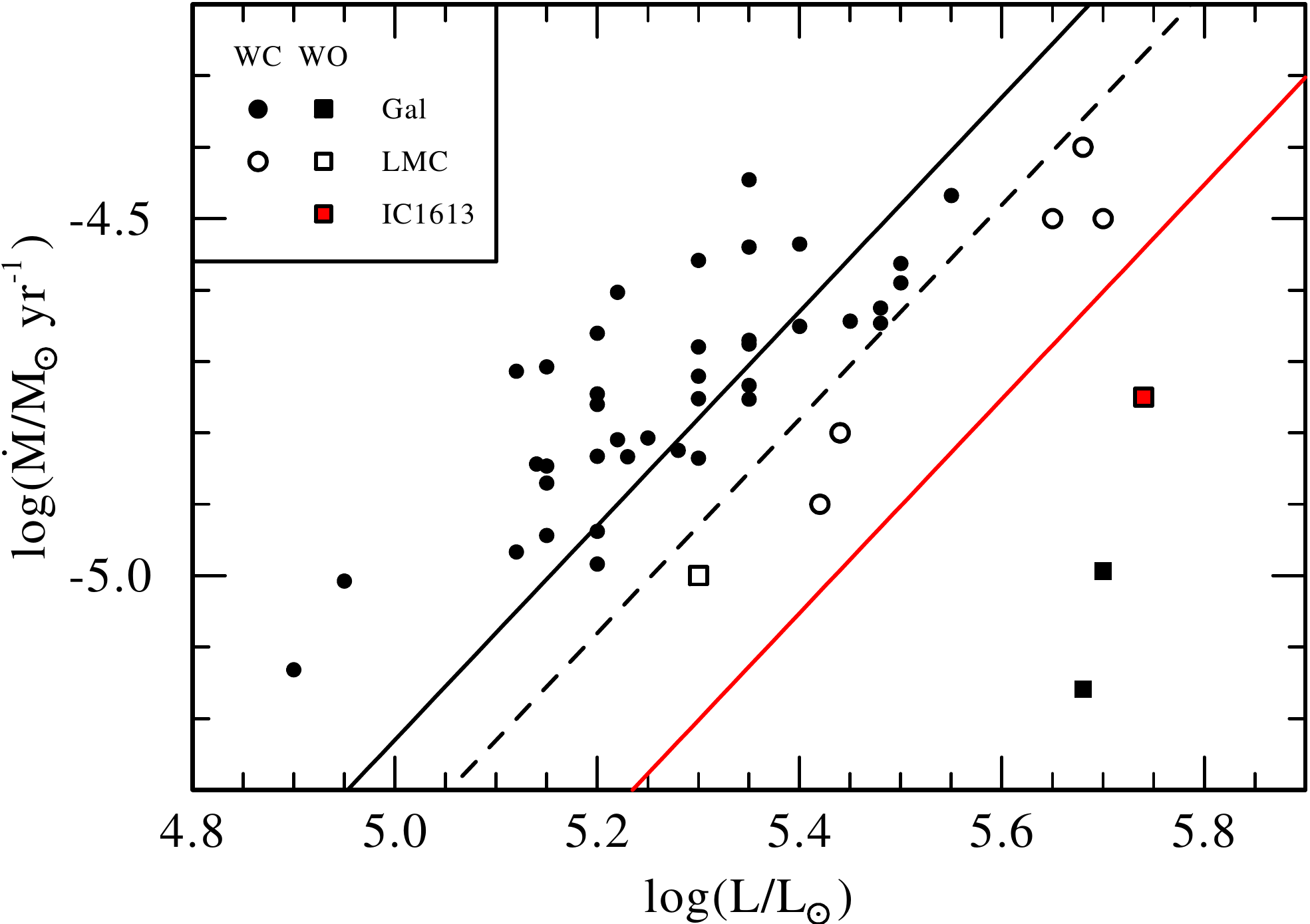}}
  \caption{Comparison of the mass-loss rates and luminosity of WC and WO stars in different metallicity environments to the mass-loss relation WR1 of \cite{yoon2005}. The solid black line represents the relation for $Z = Z_{\odot}$, the dashed black line $Z=1/2\, Z_{\odot}$, and the solid red line $Z = 1/7\, Z_{\odot}$. The solid symbols represent the results from \cite{sander2012}, the open symbols represent the results from \cite{crowther2000, crowther2002}, and the red square indicates the location of DR1.}
  \label{fig:mdotL}
\end{figure}

 Figure~\ref{fig:mdotL} shows a comparison of the mass-loss rate and luminosity of DR1 to the observed values for WC and WO stars in the Galaxy \citep{sander2012} and the LMC \citep{crowther2000, crowther2002}. The plot clearly shows that the mass-loss rate of WC stars depends on stellar luminosity and the metallicity of the environment. This is in line with model predictions by \cite{vink2005} and \cite{grafener2008}, who find that the metallicity dependence of WR mass loss is mainly controlled by the iron abundance.
\par DR1 fits in this picture, as its mass-loss rate is well below the values found for WC stars of similar luminosity in the LMC. The two Galactic WO stars studied by \cite{sander2012} have surprisingly low mass-loss rates. This could be caused by the modeling of the \ion{O}{vi} $\lambda\lambda$3811-34 line, as this requires a low mass-loss rate to prevent the emission from being re-absorbed in the stellar wind. The mass-loss rate of the LMC WO star analysed by \cite{crowther2000}, who do not fit the \ion{O}{vi} $\lambda\lambda$3811-34 emission, is consistent with the metallicity trend.
\par Alternatively, the low mass-loss rates of the Galactic WO stars may be an indication that these stars do not follow a well-defined dependence on $L$ and $Z$. One of the reasons for this could be their high temperature. Temperature effects for WR stars are predicted by \cite{grafener2008}, albeit in a different temperature regime. As pointed out by \cite{grafener2012}, there may exist a dichotomy between the winds of early WR subtypes which are likely driven by the hot iron opacity peak at $~150$\,kK, and later subtypes which may have radially inflated envelopes and much cooler winds \citep[cf. also][]{nugis2002, lamers2002}. The existence of such inflated envelopes has been questioned by \cite{petrovic2006}. 
\par Figure~\ref{fig:mdotL} also shows scaled-down fits to empirical mass-loss rates by \cite{hamann1995}, as provided by \citet[][their relation WR1]{yoon2005}.  The scaling reduces the empirical rates by a factor of six to correct for the effect of clumping.  The mass-loss rate is assumed to scale proportional to $Z^{0.5}$. This prescription reproduces the observed luminosity and metallicity dependence of WR mass loss reasonably well, save for the two Galactic WO stars.

\subsection{Ionizing flux \& nebular properties}
The observed nebular properties indicate that the nebula surrounding DR1 is one of the most extreme \ion{He}{ii} emitting regions known. Nevertheless, the properties can be well modelled with DR1 as the dominating ionizing source. Our best stellar model has an ionizing flux ratio of $\log(Q_2/Q_0) = -1.5$. This is much lower than the values ranging from $-$0.3 to $-$0.9 found in hot WC models \citep{smith2002}. These models predict that nebular \ion{He}{ii} emission is likely not observable for these values of $Q_0$ and $Q_2$, based on a predicted line strength of \ion{He}{ii} $\lambda$4686/\ion{H}{$\beta$} = $2.14 \ Q_2/Q_0$ \citep{schaerer1998}. Even though our value of $\log(Q_2/Q_0)$ is much lower \citep[a line ratio of \ion{He}{ii} $\lambda$4686/\ion{H}{$\beta$} = 0.07 is predicted by][]{schaerer1998}, we can reproduce the observed \ion{He}{ii} emission by adopting a non-homogeneous density profile. The \ion{He}{ii} emission is also stronger than predicted because the electron temperature of DR1's surrounding nebula is much higher than used in the predictions ($T_{\mathrm{e}}$ = 17.5 kK versus 10 kK).

\subsection{Final fate}
The eventual fate of DR1 is mostly determined by its mass prior to supernova explosion. Stars with a final mass $> 10 \ M_{\odot}$ are likely to form black holes \citep[e.g.,][]{georgy2009}, producing a faint supernova or no supernova at all. If rapidly rotating, however, it is possible that such massive stars produce a bright type Ib/c supernova, possibly with an associated gamma-ray burst \citep[e.g.,][]{woosley2006}. This scenario is more likely at low metallicities, as the mass-loss rates are lower during the various stages of evolution, reducing the angular momentum loss. The massive core can therefore more easily retain the angular momentum needed to power the explosion.
\par \cite{sander2012} suggest that the two Galactic WO stars in their analysis have high rotational velocities ($\approx 1000 \ \mathrm{km \ s}^{-1}$) based on the shape of the lines. If this is indeed the case, these stars would be potential progenitors of GRBs. Even though the line shapes of DR1 can be well fitted by a non-rotational model, we cannot exclude that the star is fairly rapidly rotating, as convolving with rotational profiles of up to $500 \ \mathrm{km \ s}^{-1}$ ($\approx 0.25 \, v_{\mathrm{crit}}$) has a negligible effect on the line shapes. Larger projected rotational velocities broaden the lines too much to fit our data and can thus be excluded.
\par Despite the efforts of finding the progenitors of type Ib/c SNe in pre-supernova images, none have been identified so far \citep[e.g.,][]{eldridge2013}. If the progenitors are hot WR stars like DR1, they would be very hard to detect, as despite their high bolometric luminosity, their visual brightness is very low due to their very high temperature \citep{yoon2012}.
\par Although there is still a considerable amount of helium present in our DR1 model, this does not exclude a final explosion in the form of a type Ic supernova. In single stars the helium mass fraction at the surface can be as large as 50 per cent without helium being detected in the spectrum of the supernova \citep{dessart2010}. The surface helium mass fraction of DR1 is below that level (44 per cent).

\section{Summary}\label{sec:conclusions}
In this paper we have presented a quantitative spectroscopic analysis of DR1, one of only two WO stars known at a SMC-like metallicity. We have modeled the X-Shooter spectrum using {\sc cmfgen} in order to derive the stellar and wind parameters. Our best-fit model reproduces the strength and shape of all the He, O and C lines in the 320-2000~nm wavelength range covered by our observations, including the \ion{O}{vi} $\lambda$5290 line and the  \ion{O}{vi} $\lambda$5290 / \ion{O}{v} $\lambda$5592 ratio. However, our model cannot  reproduce  the strong \ion{O}{vi} $\lambda\lambda$3811-34,  which is the prime observational diagnostic of the WO spectral type, simultaneously with the rest of the DR1 spectrum.

\par We discussed the possibility that a significant part of the \ion{O}{vi} $\lambda\lambda$3811-34 line flux is originating from a mechanism that is not included in the modeling, for instance X-ray emission produced at (or close to) the base of the wind.  Compared to early WC stars, the stellar temperature of DR1 is high, but the surface oxygen abundance is \textit{not} enhanced. This suggests that the presence of the highly ionized oxygen emission is caused by excitation effects due to the high temperature. 

\par DR1 is embedded in the hottest known \ion{He}{ii} emitting nebula in the Local Group. The properties of the nebula are consistent with DR1 being the central ionizing source. In particular the high electron temperature of the nebula is suggestive of a very hot central source and a low ambient metallicity. The capacity of DR1 to fully ionize helium may also be relevant for our understanding of the epoch of \ion{He}{ii} reionization, believed to have occured at redshifts $z \approx 4 - 2.7$ \citep[e.g.,][]{syphers2013}. Although usually attributed to quasars, WO stars like DR1 may also have contributed to the \ion{He}{ii} ionization.

\par Our best fit model and the derived oxygen and carbon abundances suggest that DR1 should be seen as a hot (i.e.\ earlier-type) WC star, and that the presence of strong \ion{O}{vi} $\lambda\lambda$3811-34 emission in WO spectra does not necessarily imply a larger oxygen abundance, hence a more advanced evolutionary stage. This statement is of importance when comparing with evolutionary computations: WO as a spectral type -- i.e.\ defined by the presence of strong \ion{O}{vi} $\lambda$$\lambda$3811-34 emission -- is not equivalent to the definition of WO stars usually adopted from a theoretical point of view. The latter is based on an enhanced oxygen content \citep[surface abundances $(\mathrm{C} + \mathrm{O}) / \mathrm{He} > 1$ by number, e.g. ][]{smith1991}.

\par DR1 is located in the high-temperature and high-luminosity domain of the HRD. Compared to evolutionary tracks, its location is compatible with the late stages of evolution of stars with an initial mass $> 60 \ M_{\odot}$, although this number is subject to various assumptions in the theoretical tracks. The carbon and oxygen abundances and stellar temperature suggest that DR1 is currently well into its core-helium burning stage. Based on the derived mass-loss rate, the clumping of the stellar wind should be moderate ($f_c \ga 0.2$), as otherwise the star should have already exploded.

\par Although we find that the WO phase likely does not represent a stage of enhanced oxygen abundance compared to WC stars, we do not exclude the possibility that WO stars are in a more advanced evolutionary stage than WC-type Wolf-Rayet stars. The higher temperatures of WO's may be the result of progressive stripping of the outer layers through the radiation driven wind, exposing consecutively hotter layers.  

\par Alternatively, WO and WC stars may be the end products of stars from different initial mass ranges, in which case the question of evolutionary connection between WO and WC stars does not apply. Detailed spectroscopic analyses of more WO and early-WC stars are needed to decide whether the properties of DR1, in particular its high temperature and WC-like oxygen abundance,  are representative of the WO stars as a class and to investigate further the nature of the WO's and their possible evolutionary connection with WC stars.

\begin{acknowledgements}
S.d.M. acknowledges support through a Hubble Fellowship grant HST-HF-51270.01-A awarded by the STScI, operated by AURA, Inc., under contract NAS 5-26555 and a Einstein Fellowship grant PF3-140105 awarded by the Chandra X-ray Center, operated SAO under the contract NAS8-03060.
\end{acknowledgements}

\bibliographystyle{aa}
\bibliography{DR1}

\Online
\begin{appendix} 
\section{Model atoms}

\begin{table}
\centering
\caption{Overview of the model atoms used.}\label{tab:species}
\begin{tabular}{l c c c}
\hline\hline
Species & $N_i$ & $N_s$ & $N_t$\\
\hline \\[-8pt]
\ion{He}{i} & 27 & 27 & 27 \\
\ion{He}{ii} & 13 & 13 & 30 \\
\ion{C}{ii} & 22 & 22 & 22 \\
\ion{C}{iii} & 44 & 100 & 243 \\ 
\ion{C}{iv} & 59 & 59 & 64 \\
\ion{O}{ii} & 3 & 3 & 3 \\
\ion{O}{iii} & 79 & 79 & 115 \\
\ion{O}{iv} & 53 & 53 & 72 \\
\ion{O}{v} & 75 & 75 & 152 \\
\ion{O}{vi} & 25 & 25 & 31 \\
\ion{Ne}{ii} & 0 & 42 & 242 \\
\ion{Ne}{iii} & 10 & 40 & 182 \\
\ion{Ne}{iv} & 10 & 45 & 355 \\
\ion{Ne}{v} & 10 & 37 & 166 \\
\ion{Ne}{vi} & 10 & 36 & 202 \\
\ion{Ne}{vii} & 10 & 38 & 182 \\
\ion{Ne}{viii} & 24 & 24 & 47 \\
\ion{Si}{iv} & 10 & 37 & 48 \\
\ion{Si}{v} & 10 & 33 & 71 \\
\ion{Si}{vi} & 20 & 42 & 132 \\
\ion{P}{iv} & 0 & 36 & 178 \\
\ion{P}{v} & 0 & 16 & 62 \\
\ion{S}{iii} & 0 & 13 & 28 \\
\ion{S}{iv} & 0 & 51 & 142 \\
\ion{S}{v} & 0 & 31 & 98 \\
\ion{S}{vi} & 28 & 28 & 58 \\
\ion{Cl}{iv} & 0 & 40 & 129 \\
\ion{Cl}{v} & 0 & 26 & 80 \\
\ion{Cl}{vi} & 0 & 18 & 44 \\
\ion{Cl}{vii} & 0 & 17 & 28 \\
\ion{Ar}{iii} & 0 & 32 & 346 \\
\ion{Ar}{iv} & 0 & 50 & 382 \\
\ion{Ar}{v} & 0 & 64 & 376 \\
\ion{Ar}{vi} & 0 & 21 & 81 \\
\ion{Ar}{vii} & 0 & 30 & 72 \\
\ion{Ar}{viii} & 0 & 28 & 52 \\
\ion{Ca}{iii} & 0 & 33 & 110 \\
\ion{Ca}{iv} & 0 & 34 & 193 \\
\ion{Ca}{v} & 0 & 45 & 121 \\
\ion{Ca}{vi} & 0 & 47 & 108 \\
\ion{Ca}{vii} & 0 & 48 & 288 \\
\ion{Ca}{viii} & 0 & 45 & 296 \\
\ion{Ca}{ix} & 0 & 39 & 162 \\
\ion{Ca}{x} & 27 & 27 & 59 \\
\ion{Cr}{iv} & 0 & 29 & 234 \\
\ion{Cr}{v} & 0 & 30 & 223 \\
\ion{Cr}{vi} & 0 & 30 & 215 \\
\ion{Mn}{iv} & 0 & 39 & 464 \\
\ion{Mn}{v} & 0 & 16 & 80 \\
\ion{Mn}{vi} & 0 & 23 & 181 \\
\ion{Mn}{vii} & 0 & 20 & 203 \\
\ion{Fe}{iv} & 51 & 51 & 294 \\
\ion{Fe}{v} & 47 & 47 & 191 \\
\ion{Fe}{vi} & 44 & 44 & 433 \\
\ion{Fe}{vii} & 41 & 41 & 252 \\
\ion{Fe}{viii} & 53 & 53 & 324 \\
\ion{Fe}{ix} & 52 & 52 & 490 \\
\ion{Fe}{x} & 43 & 43 & 210 \\
\ion{Ni}{iv} & 0 & 36 & 200 \\
\ion{Ni}{v} & 0 & 46 & 183 \\
\ion{Ni}{vi} & 0 & 37 & 314 \\
\ion{Ni}{vii} & 0 & 37 & 308 \\
\ion{Ni}{viii} & 0 & 34 & 325 \\
\ion{Ni}{ix} & 0 & 34 & 363 \\
\hline
\end{tabular}
\tablefoot{
$N_i$ is the number of levels that are treated with an accelerated lambda iteration. $N_s$ is the number of superlevels, each of which may consist of a single level or multiple levels. $N_t$ the total number of atomic levels in the model atom.
}
\end{table}

\end{appendix}
\end{document}